\documentclass[]{aa}
\usepackage{graphicx,url,threeparttable,subfig,booktabs,placeins}  \usepackage[version=4]{mhchem}
\usepackage{mysiunitx}
\usepackage{derivative}

\usepackage[pdftex,
pdfauthor={Matthias Fabry, Pablo Marchant, Norbert Langer, Hugues Sana},
pdftitle={Modeling contact binaries, III. Properties of a population of close, massive binaries},
pdfkeywords={Stars: massive, Binaries: Close},
pdfproducer={Latex with hyperref},
pdfcreator={pdflatex}]{hyperref}
\hypersetup{colorlinks=true,citecolor=blue}

\bibpunct{(}{)}{,}{a}{}{,}

\DeclareSIUnit{\year}{yr}
\DeclareSIUnit{\Msun}{\mathnormal{M_\odot}}
\DeclareSIUnit{\Lsun}{\mathnormal{L_\odot}}
\DeclareSIUnit{\Rsun}{\mathnormal{R_\odot}}
\DeclareSIUnit{\kms}{\kilo\meter\per\second}
\newcommand{\unsim}{\mathord{\sim}}
\newcommand{\Lag}[1]{\ensuremath{{\rm L}_{#1}}}

\defcitealias{endalEvolutionRotatingStars1976}{ES}
\defcitealias{coxPrinciplesStellarStructure1968}{CG}
\defcitealias{fabryModelingOvercontactBinaries2022}{Paper I}
\defcitealias{fabryModelingContactBinaries2023}{Paper II}
\defcitealias{shuStructureContactBinaries1976}{SLA}

\begin{document}

\title{Modeling contact binaries}
\subtitle{III. Properties of a population of close, massive binaries}
\titlerunning{Modeling contact binaries, III. Population of massive binaries}

\author{
M. Fabry\inst{\ref{vu}, \ref{ivs}} 
\and P. Marchant\inst{\ref{ivs}}
\and N. Langer\inst{\ref{aifa}, \ref{mpifr}}
\and  H. Sana\inst{\ref{ivs}, \ref{lgi}}
}

\institute{
Department of Astrophysics and Planetary Science, Villanova University, 800 E Lancaster Ave., Villanvona, PA 19085 \\ \email{matthias.fabry@villanova.edu} \label{vu}
\and
Institute of Astronomy, KU Leuven, Celestijnenlaan 200D, 3001 Leuven, Belgium \label{ivs}
\and
Leuven Gravity Institute, KU Leuven, Celestijnenlaan 200D box 2415, 3001 Leuven, Belgium\label{lgi}
\and
Argelander-Institut f\"ur Astronomie, Universit\"at Bonn, Auf dem H\"ugel 71, D-53121 Bonn, Germany \label{aifa}
\and
Max-Planck-Institut f\"ur Radioastronomie, Auf dem Hügel 69, D-53121 Bonn \label{mpifr}
}

\abstract
{
Among massive stars, binary interaction is the rule rather than the exception.
The closest binaries, those with periods of less than $\unsim$10 days, undergo mass transfer during core-hydrogen burning, with many of them experiencing a nuclear-timescale contact phase.
Current binary population synthesis models predict the mass-ratio distribution of contact binaries to be heavily skewed toward a mass ratio of unity, which is inconsistent with observations.
It has been shown that effects of tidal deformation due to the Roche potential, as well as energy transfer in the common layers of a contact binary, alter the internal structure of close binary components.
However, previous population studies neglected these effects.
}
{
We model a population of massive binary stars that undergo mass transfer during core-hydrogen burning, while consistently considering the effects of tidal deformation and energy transfer in contact phases.
}
{
We use the MESA binary-evolution code to compute large grids of models with primary star masses of $\qtyrange{8}{70}{\Msun}$ at Solar metallicity.
We then perform a population synthesis study to predict distribution functions of the observational properties of close binary systems, focusing in particular on the mass and luminosity ratio distribution.
}
{
We find that the effects of tidal deformation and energy transfer have a limited effect on the predicted mass-ratio distribution of massive contact binaries.
Only a small fraction of the population has their mass ratio significantly shifted toward a more unequal configuration.
However, we suggest that orbital hardening could affect the evolution of contact binaries and their progenitors, and we advocate for a homogeneous set of observed contact binary parameters.
}
{}


\maketitle

\section{Introduction}
Among massive-star ($M_{\rm init} \gtrsim \SI{8}{\Msun}$) populations, the binary fraction is observed to be very high, and the majority of stars are born in systems that will undergo strong interactions \citep{masonHighAngularResolution2009, sanaBinaryInteractionDominates2012, sanaVLTFLAMESTarantulaSurvey2013, sanaSouthernMassiveStars2014, kobulnickyCOMPLETESTATISTICSMASSIVE2014, dunstallVLTFLAMESTarantulaSurvey2015, guoBinarityEarlytypeStars2022, lanthermannMultiplicityNorthernBright2023}.
The observed period distribution of these binaries is consistent with being (logarithmically) flat, $\mathcal{N}(\log p) \sim (\log p)^{0}$, over periods ranging from one to several hundred days, meaning a significant fraction of binaries have short initial periods ($p_{\rm init} \lesssim \qty{10}{\day}$, \citealp{sanaVLTFLAMESTarantulaSurvey2013, almeidaTarantulaMassiveBinary2017,villasenorBtypeBinariesCharacterization2021, banyardObservedMultiplicityProperties2022}). 
Many population synthesis studies show such binaries interact on the main sequence, following so-called Case A evolution \citep[e.g.,][]{polsCaseEvolutionMassive1994, vanbeverenWROtypeStar1998, nelsonCompleteSurveyCase2001, deminkRotationRatesMassive2013, senDetailedModelsInteracting2022}.
\par

Accurately modeling binary interaction is of paramount importance to understand the evolution of massive stars \citep{marchantEvolutionMassiveBinary2024}.
Close binary interaction leads to phenomena that cannot be explained by single-star models, such as stellar mergers and their associated magnetism \citep{schneiderStellarMergersOrigin2019, frostMagneticMassiveStar2024}, super-luminous supernovae \citep{yoonEvolutionRapidlyRotating2005, yoonSingleStarProgenitors2006, aguilera-denaRelatedProgenitorModels2018, aguilera-denaPrecollapsePropertiesSuperluminous2020}, and results in directly observable interaction products, such as stripped stars \citep[e.g.,][]{shenarHiddenCompanionLB12020, bodensteinerDetectingStrippedStars2022, droutObservedPopulationIntermediatemass2023} and short-period Wolf-Rayet binaries \citep{hainichWolfRayetStarsLarge2014,pauliSyntheticPopulationWolfRayet2022}.
Furthermore, close massive binary evolution can yield gravitational-wave merger events through various channels.
In particular, the chemically homogeneous evolution channel \citep{marchantNewRouteMerging2016, mandelMergingBinaryBlack2016} favors very close binaries, those with periods of around $\qty{1}{\day}$, which in most cases means such binaries experience extended periods of contact evolution.
\par

In very close binary systems ($p \lesssim \qty{2}{\day}$), a contact phase is likely to occur \citep{wellsteinFormationContactMassive2001, menonDetailedEvolutionaryModels2021, hennecoContactTracingBinary2024}.
Such phases are characterized by both stars overflowing their respective Roche lobes (RLs) simultaneously.
In this configuration, the outer layers of the stars are heavily deformed by the tidal interaction between the companions.
Additionally, there is thermal contact in the common layers of a contact binary, which allows for the transfer of energy between the components just as the mechanical contact allows for mass transfer.
\par

Hundreds to thousands of low-mass contact binaries, called W Ursae Majoris (W UMa) stars, have been discovered thanks to both targeted and all-sky photometric surveys \citep{eggenContactBinariesII1967, binnendijkOrbitalElementsUrsae1970, szymanskiContactBinariesOGLEI2001, selamKeyParametersUMatype2004, paczynskiEclipsingBinariesASAS2006, graczykOpticalGravitationalLensing2011, graczykLatetypeEclipsingBinaries2018}.
Conversely, only a few dozen contact systems with massive components are known, resulting from the MACHO survey \citep{alcockMACHOProjectLMC1997}, the Tarantula Massive Binary Monitoring \citep[TMBM,][]{almeidaDiscoveryMassiveOvercontact2015,almeidaTarantulaMassiveBinary2017, mahyTarantulaMassiveBinary2020b,mahyTarantulaMassiveBinary2020a} and the analysis of several galactic objects \citep[e.g.,][]{polushinaCatalogueMassiveClose2004, pennyTomographicSeparationComposite2008, lorenzoMYCamelopardalisVery2014, lorenzoMassiveMultipleSystem2017, yangComprehensiveStudyThree2019, abdul-masihConstrainingOvercontactPhase2021, mahyTarantulaMassiveBinary2020b, liV606CenNewly2022}.
Contact binary light curves are characterized by smooth, near-sinusoidal variability, without flattened (partial) eclipses, signaling that the components are of comparable size and temperature.
Using a Wilson-Devinney-like method \citep[][and later updates]{wilsonRealizationAccurateCloseBinary1971}, the analysis of the light curve allows, among others, to determine the degree of overflow, $(R - R_{\rm RL})/R_{\rm RL}$, of the components.
Unfortunately, this value is rather uncertain from the light curve alone, and it is sometimes unclear whether a system is in true contact, or only in near-contact \citep[e.g.,][]{hilditchFortyEclipsingBinaries2005,mahyTarantulaMassiveBinary2020b}.
A spectroscopic characterization can be performed to get dynamical masses and thus an independent constraint on the mass ratio as well as the projected physical separation.
The combination of these analyses, especially when accounting for the tidally distorted nature of the photosphere, provides the best measurements of contact binary parameters \citep{abdul-masihSpectroscopicPatchModel2020, abdul-masihConstrainingOvercontactPhase2021}.
\par

In the literature, most efforts to theoretically model contact binaries are focused on the W UMa stars.
Many models included a prescription of energy transfer \citep[e.g.,][]{lucyStructureContactBinaries1968, biermannModelsContactBinaries1972, flanneryCyclicThermalInstability1976, shuStructureContactBinaries1976, kahlerStructureEquationsContact1989}, but some turned out to either produce inaccurate light curves or contain inconsistencies.
Unfortunately, models that are applicable to massive stars with radiative envelopes are more limited.
\citet{shuStructureContactBinaries1976} provided the contact-discontinuity model, but it is challenged by \citet{papaloizouMaintenanceTemperatureDiscontinuity1979}, \citet{hazlehurstStabilityAgezeroContact1980} and \citet{hazlehurstEquilibriumContactBinary1993}.
Given the problems of developing a complete theory of contact binary structure, many recent studies that focused on close massive binary evolution and included contact phases, like \citet{deminkRotationRatesMassive2013}, \citet{marchantNewRouteMerging2016}, \citet{menonDetailedEvolutionaryModels2021}, \citet{senDetailedModelsInteracting2022} and \citet{hennecoContactTracingBinary2024} ignored energy transfer in the common layers of a contact system.
The models find that contact phases can occur in both thermally unstable and thermally stable evolution.
In the former, a rapid mass-transfer phase causes the accretor to swell beyond its RL as it falls out of thermal equilibrium, while in the latter, the evolution is dominated by the nuclear timescale of the system.
However, these state-of-the-art models have trouble reproducing the observed mass-ratio distribution of contact systems.
While they indicate that, once nuclear-timescale contact is engaged, the binary should equalize rapidly \citep{menonDetailedEvolutionaryModels2021}, long-term monitoring of the period derivative shows that the orbit evolves on the nuclear timescale \citep{abdul-masihConstrainingOvercontactPhase2022, vranckenConstrainingOvercontactPhase2024}.
However, \citet{fabryModelingContactBinaries2023} showed that for particular systems, energy transfer in radiative envelopes can alter the mass-ratio evolution of massive contact systems, and has the possibility to relieve part of the discrepancy between models and observations.
\par

In this paper, we use the methodology developed in \citet{fabryModelingOvercontactBinaries2022} and \citet{fabryModelingContactBinaries2023}, henceforth \citetalias{fabryModelingOvercontactBinaries2022} and \citetalias{fabryModelingContactBinaries2023}, respectively, to consistently take into account tidal deformation of close binary components and energy transfer in contact phases.
We compute a grid of short-period, massive binary-evolution models and investigate the effect of energy transfer on the mass-ratio distribution of contact systems.
The paper is organized as follows.
Section \ref{sec:meth} describes the setup of the stellar-evolution calculations, the assumptions on stellar and binary physics, the initial and termination conditions, and we specify the explored parameter space to provide population predictions.
Section \ref{sec:results} presents and discusses the results obtained from the population-synthesis modeling, while Sect. \ref{sec:conc} gives concluding remarks.
 \section{Methods}\label{sec:meth}

We use the detailed binary-evolution code \texttt{MESA}, version 22.11.1 \citep{paxtonModulesExperimentsStellar2011, paxtonModulesExperimentsStellar2013, paxtonModulesExperimentsStellar2015, paxtonModulesExperimentsStellar2018, paxtonModulesExperimentsStellar2019, jermynModulesExperimentsStellar2023}, to compute grids of models that undergo Case A mass transfer, from which we assemble a synthetic population. All models are computed at solar metallicity, $Z_\odot = 0.0142$, with metal fractions following \citet{asplundChemicalCompositionSun2009}.
The parameters to be varied in each grid are the initial primary mass, $M_{1, \rm init}$, initial mass ratio, $q_{\rm init} = M_{2,{\rm init}}/M_{1,\rm init}$, and the initial period, $p_{\rm init}$.
For initial primary mass, we sample $M_{1, \rm init} = \{8, 9, 10, ..., 19\} \cup \{20, 22, 24, ..., 48\} \cup \{50, 52.5, 55, ..., 70\} \unit{\Msun}$, which is chosen to cover the mass range of all observed (near-)contact binaries (accounting for significant mass-loss of the most massive stars).
The initial mass ratio is $q_{\rm init} = \numrange{0.6}{0.975}$, spaced uniformly with $\Delta q_{\rm init} = 0.025$, while the initial period is $p_{\rm init} = \qtyrange{0.5}{8}{\day}$, spaced logarithmically with $\Delta\log(p/\unit{\day}) = 0.04$.
We compute three grids with the following parameter variations: one that includes the effect of energy transfer (ET) and tidal deformation, one without ET but with tidal deformation, and the last without ET and using the single-rotating star deformation corrections from \citet{paxtonModulesExperimentsStellar2019}.
This amounts to a total of \num{53568} detailed binary-evolution models.
We require these extensive grids to precisely identify where ET has the greatest impact in the parameter space.
Throughout the discussion, we will refer to the models that include ET and tidal deformation as the ``ET models,'' those that do not (but still use the tidal deformation) the ``no-ET models,'' and those that use the single-rotating star deformation the ``single-rotating models.''

\subsection{Physical assumptions in \texttt{MESA}}\label{ssec:mesa_assump}
We use a similar setup of the \texttt{MESA} code as in \citetalias{fabryModelingContactBinaries2023}, that is, we use the same microphysics, convection and overshoot model, wind prescription, deformation geometry, and mass- and energy-transfer calculations.
We present briefly the general setup, while the significant changes with respect to \citetalias{fabryModelingOvercontactBinaries2022} and \citetalias{fabryModelingContactBinaries2023} are listed at the end of this subsection.
\par

In terms of microphysics, we use the OPAL radiative opacities from \citet{iglesiasRadiativeOpacitiesCarbon1993, iglesiasUpdatedOpalOpacities1996} and \citet{fergusonLowTemperatureOpacities2005} for low-temperature opacities.
For high-temperatures, we use the Compton scattering opacities from \citet{poutanenRosselandFluxMean2017} and electron conduction opacities from \citet{cassisiUpdatedElectronconductionOpacities2007} and \citet{blouinNewConductiveOpacities2020}.
The equation of state is a blend of \citet{saumonEquationStateLowMass1995}, \citet{timmesAccuracyConsistencySpeed2000}, \citet{rogersUpdatedExpandedOPAL2002}, \citet{potekhinThermodynamicFunctionsDense2010} and \citet{jermynSkyeDifferentiableEquation2021}, specified in \citet{jermynModulesExperimentsStellar2023} (except that we exclude the FreeEOS table).
We use the basic nuclear-reaction network involving the species $\ce{^1H}, \ce{^3He}, \ce{^4He}, \ce{^{12}C}, \ce{^{14}N}, \ce{^{16}O}, \ce{^{20}Ne}$ and $\ce{^{24}Mg}$, with reaction rates taken from the JINA \citep{cyburtJINAREACLIBDatabase2010} and NACRE \citep{anguloCompilationChargedparticleInduced1999} libraries, with weak interaction tables from \citet{fullerStellarWeakInteraction1985}, \citet{odaRateTablesWeak1994} and \citet{langankeShellmodelCalculationsStellar2000}. Electron screening is computed as per \citet{chugunovCoulombTunnelingFusion2007}, and thermal neutrino losses are taken from \citet{itohNeutrinoEnergyLoss1996}.
\par

Wind-mass loss is computed as per \citet{brottRotatingMassiveMainsequence2011}.
If the surface hydrogen fraction is $X > 0.7$, the mass-loss rate is \citet{vinkMasslossPredictionsStars2001} if the temperature is above the iron bi-stability jump (also calibrated by \citet{vinkMasslossPredictionsStars2001}, while below that temperature, the maximum of the rates of \citet{vinkMasslossPredictionsStars2001} and \citet{nieuwenhuijzenAtmosphericAccelerationsStability1995} is chosen.
For surface hydrogen fractions $X < 0.4$, the mass-loss prescription from \citet{hamannSpectralAnalysesGalactic1995} is used, reduced by a factor of 10.
Between $0.4 < X < 0.7$, the mass-loss rate is linearly interpolated between the above schemes.
Wind mass transfer is accounted for through the Bondi-Hoyle mechanism \citep{bondiMechanismAccretionStars1944}, as implemented by \citet{hurleyEvolutionBinaryStars2002}.
\par

Convective regions are determined following the Ledoux criterion \citep{ledouxStellarModelsConvection1947}, where the mixing length theory of \citet{vitenseWasserstoffkonvektionszoneSonneMit1953} and \citet{bohm-vitenseUberWasserstoffkonvektionszoneSternen1958} is applied (as described by \citealp{coxPrinciplesStellarStructure1968}).
Step overshooting is used as calibrated by \citet{brottRotatingMassiveMainsequence2011}, where the diffusion coefficient from $0.01$ pressure scale heights inside the boundary is kept constant to $0.335$ pressure scale heights outside the convective-core boundary.
We account for the semiconvective instability as treated by \citet{langerSemiconvectiveDiffusionEnergy1983}, and thermohaline mixing from \citet{kippenhahnTimeScaleThermohaline1980}, both with an efficiency of unity, $\alpha_{\rm sc} = \alpha_{\rm th} = 1$.
\par

Mass transfer (MT) in semi-detached configurations is computed implicitly by requiring for the donor to stay just below its RL radius (as approximated by \citealp{eggletonApproximationsRadiiRoche1983}).
For contact configurations, the MT rate is adjusted so that the radii of the stars share an equipotential surface, amounting to a relation
\begin{equation}
    r_{2} = F(q, r_{1}),
\end{equation}
where $r = (R - R_{\rm RL})/R_{\rm RL}$ is the relative RL overflow of a component.
Within the grids with the tidal deformation corrections, we use the integrations of \citetalias{fabryModelingOvercontactBinaries2022} to construct $F$, while in the case of the single-rotating star corrections, we use $F(q, x) = q^{-0.52}x$, as in \citet{marchantNewRouteMerging2016}.
ET is computed like in \citetalias{fabryModelingContactBinaries2023}, by assuming there is an efficient energy flow from one component to the other in the layers near the RL equipotential.
Energy is redistributed so that each component radiates a fraction of the total luminosity $L_1 + L_2$ proportional to its surface area $S$ and surface averaged effective gravity $\langle g\rangle$, as we assumed von Zeipel's theorem of radiative equilibrium holds \citep{vonzeipelRadiativeEquilibriumRotating1924}.
The combination of using the tidal deformation corrections and efficient ET ensures shellularity of the shared envelope of a contact-binary model \citepalias{fabryModelingContactBinaries2023}.
\par

With respect to \citetalias{fabryModelingOvercontactBinaries2022} and \citetalias{fabryModelingContactBinaries2023}, we make the following modifications to the modeling setup:
\begin{itemize}
	\item We do not assume rigid-body rotation throughout the evolution, so that differential rotation is allowed.
	However, since we expect fast tidal synchronization for very close binaries \citep{zahnTidalFrictionClose1977}, we uniformly modify the angular momentum of the star on the timescale of the orbital period, $p$,
	\begin{equation}\label{eq:population:tides}
		\Delta j_{\rm tides} = \left(1 - \exp\left(\frac{-\Delta t}{p}\right)\right)\left(\omega_{\rm orb} i_{\rm rot} - j\right).
	\end{equation}
	Here $\Delta t$ is the timestep of the evolution, $j$ and $i_{\rm rot}$ are the current angular momentum and the moment of inertia of the stellar layer, respectively, and $\omega_{\rm orb} = 2\pi/p$ is the orbital angular velocity.
	We model spin-orbit coupling to conserve total angular momentum, so that the required angular momentum to synchronize both stars in this manner is then subtracted from the orbital angular momentum.
	Given that $\Delta t$ is much longer than the orbital period in evolutionary calculations (years to thousands of years versus days), Eq.~\eqref{eq:population:tides} specifies efficient tidal synchronization.
	We thus expect little departure from solid-body rotation in our very close binary simulations, since processes that cause departures from it operate on longer timescales than the orbital one.

	\item We also allow for rotational mixing in our models.
	We include Eddington-Sweet circulation \citep{eddingtonCirculatingCurrentsRotating1925, sweetImportanceRotationStellar1950}, the Goldreich-Schubert-Fricke instability \citep{goldreichDifferentialRotationStars1967, frickeInstabilitatStationarerRotation1968} and both the dynamical- and secular-shear instability \citep{zahnCirculationTurbulenceRotating1992, endalEvolutionRotatingStars1978}, all following the implementation of \citet{hegerPresupernovaEvolutionRotating2000}.
	
	\item We reduce the parameters of mixing length theory to $\alpha_{\rm MLT} = 1.5$, and the semiconvective efficiency to $\alpha_{\rm sc} = 1$, to be in line with \citet{menonDetailedEvolutionaryModels2021}.

	\item When using the single-rotating star deformation corrections, we setup the code to follow as closely as possible the settings used in \citet{menonDetailedEvolutionaryModels2021}.
	No ET is applied for this grid of models.
	The specific moment of inertia of a stellar shell of radius $r_\Psi$ is evaluated simply as $i_{\rm rot} = 2/3\,r_{\Psi}^2$, while the structure corrections $f_P$ and $f_T$ follow the analytical fits of \citet{paxtonModulesExperimentsStellar2019}, with the modification that the corrections are bounded to their values at 60\% critical rotational velocity.
	We emphasize that this does not mean that a model cannot rotate faster than this limit, only that the structure corrections are limited to values above 0.74 and 0.93 for $f_P$ and $f_T$, respectively.
	From the results of \citetalias{fabryModelingOvercontactBinaries2022}, the single-rotating star corrections slightly underestimate the deformation as compared to the tidal corrections.
\end{itemize}
\par

\subsection{Binary initialization}\label{ssec:binary_init}
Once initial masses and a period are selected, the binary model is initialized as follows.
Non-rotating, single-star models of mass $M_{1, \rm init}$ and $M_{2,\rm init} = qM_{1, \rm init}$ are interpolated from pre-computed zero-age main sequence (ZAMS) models of solar metallicity and a helium content of $Y=0.2703$.
We set the eccentricity to zero as we expect circular orbits for close binaries \citep{zahnDynamicalTideClose1975}.
The single-star models are then spun up to the appropriate Keplerian angular velocity $\omega_{\rm orb}$.
Since the models are brought out of thermal equilibrium due to the spin up, what follows is a thermal relaxation period onto the true ZAMS of the models, during which MT, ET or any other mass loss is turned off.
We also keep the period constant to the initial value and enforce rigid-body rotation at the orbital angular velocity.
We define the ZAMS as the point where the luminosity of both stars is within 1\% of their nuclear luminosity.
When this condition is reached, we make a decision depending on the rate of overflow of the stars.
\par

First, if both stars are within their respective RLs, the simulation proceeds as normal, without further intervention.
Second, if at least one star overflows its RL, this means that MT must have started in the pre-main sequence (PMS) phase.
We do not take these models into account in the main body of the paper, but discuss them in Appendix \ref{app:pms_interaction}.
Finally, if one of the stars overflows the second Lagrangian point (L2OF) at the ZAMS, we terminate the simulation immediately as we assume a merger would have happened on the PMS.
\par

\subsection{Outcomes and termination}\label{ssec:outcomes}
The outcome of our simulations broadly falls in three categories.
\begin{itemize}
	\item The binary reaches $\Lag{2}$ in a contact configuration.
	In this situation, we assume mass loss will start from $\Lag{2}$, carrying away a large amount of specific angular momentum \citep{marchantRoleMassTransfer2021}, which we expect results in a merger on a short timescale.
	We stop the simulation at this point and classify these systems as ``mergers.''
	A variant of this outcome is when the MT calculation determines that $\dot{M}$ is a very high value.
	We choose to take a limiting value of $\dot{M} = \qty{e-1}{\Msun\per\year}$ across the whole grid.
	MT rates higher than this value are at least two orders of magnitude above the thermal timescale MT rates of massive main-sequence stars (except perhaps for the highest masses in our grid, where $\dot{M}_{\rm thermal} \approx \qty{5e-3}{\Msun\per\year}$), and we expect MT to evolve toward dynamical timescale MT, which is associated with highly non-conservative evolution, leading to a likely merger.
	
	\item One of the stars (most likely the initially more massive one) reaches core-hydrogen exhaustion, without triggering a merger.
	At this point, the star will start undergoing rapid evolution, and we stop the simulation calling these ``(main-sequence) survivors.''
	
	\item The simulation can encounter numerical difficulties, for example by failing to reach convergent models during the evolution.
	In this case, we halt the simulation and signal this system encountered an ``error.''
\end{itemize}

\subsection{Population synthesis computations}\label{ssec:pop_synth}
The probability density, $\mathcal{P}$, of finding systems in a certain configuration $\vartheta_0$ is then proportional to the weighted contributions in that configuration,
\begin{equation}\label{eq:probability}
	\mathcal{P}(\vec{\vartheta_0}) \equiv \odv{\mathcal{N}}{\vec{\vartheta}}_{\vec{\vartheta_0}} \propto \int \delta(\vec{\vartheta} - \vec{\vartheta_0}) \mathcal{W}_\mathcal{N} \odif{\mathcal{N}},
\end{equation}
where each of the models, $\mathcal{N}$, are considered only at the configuration $\vec{\vartheta} = \vec{\vartheta_0}$, using the Dirac delta function, $\delta$.
If the required configuration is characterized by a single quantity parameterized by time, $\vartheta = \vartheta(t)$, we can simplify Eq.~\eqref{eq:probability} by transforming the delta function under composition with a function $g$:
\begin{equation}
	\delta(g(x)) = \sum_{i}\frac{\delta(x - x_i)}{\left|g'(x_i)\right|},
\end{equation}
where we sum over the (simple) roots of $g$.
Therefore,
\begin{align}
	\mathcal{P}(\vartheta) &\propto \sum_{i}\int \frac{\delta(t - t_i)}{\left|\odv{\vartheta}{t}\right|_{t_i}}\mathcal{W}_\mathcal{N}\odif{\mathcal{N}}\\
	&=\sum_{i}\mathcal{W}_\mathcal{N}(t_i)\left|\odv{\vartheta}{t}\right|_{t_i}^{-1}.
\end{align}
This shows that the probability density is inversely proportional to the velocity at which the quantity in question changes at the time where $\vartheta = \vartheta_0$ and we sum the contributions from all times $t_i$ at which $\vartheta = \vartheta_0$.
For $n$-dimensional configurations $\vec{\vartheta}$, the mathematics is not as straightforward to write down, but the result is that we should sum over the $(n-1)$-dimensional hypersurface where $\vec{\vartheta} = \vec{\vartheta_0}$, which may consist of multiple disconnected parts (just as there might be different roots in the 1D case).
\par

The probability density is weighted with factors $\mathcal{W}_\mathcal{N}$, coming from the likelihood of this model being created from the star- and binary-formation processes.
Since we vary the initial primary mass, initial mass ratio and initial period, and take into account the star-formation-rate (SFR) history, the total (relative) weight is given by,
\begin{equation}
	\mathcal{W}_\mathcal{N}\odif{\mathcal{N}} = \mathcal{W}_{M_{1, \rm init}}\odif{M_{1, \rm init}}\mathcal{W}_{q_{\rm init}}\odif{q_{\rm init}}\mathcal{W}_{p_{\rm init}}\odif{p_{\rm init}}\mathcal{W}_{\rm SFR}\odif{t}.
\end{equation}
The individual weights are taken from models of observed young populations.
Specifically, we adopt,
\begin{subequations}\label{eqs:birthdists}
	\begin{align}
		\mathcal{W}_{M_{1, \rm init}}\odif{M_{1, \rm init}} &= M_{1, \rm init}^{-2.35}\odif{M_{1, \rm init}},\\
		\mathcal{W}_{q_{\rm init}}\odif{q_{\rm init}} &= q_{\rm init}^0\odif{q_{\rm init}} = \odif{q_{\rm init}}, \\
		\mathcal{W}_{p_{\rm init}}\odif{p_{\rm init}} &= (\log p_{\rm init})^0\odif{\log p_{\rm init}},\label{eq:p-dist} \\
		\mathcal{W}_{\rm SFR}\odif{t} &= \odif{t},
	\end{align}
\end{subequations}
coming from the Salpeter initial-mass function \citep{salpeterLuminosityFunctionStellar1955, kroupaVariationInitialMass2001, bastianUniversalStellarInitial2010} and the modeled mass-ratio distributions found from a sample of O-type stars \citep{sanaBinaryInteractionDominates2012, shenarTarantulaMassiveBinary2022}.
No sample of observed binaries covers periods shorter than $\qty{1}{\day}$, so we adopt \"Opik's law \citep{opikStatisticalStudiesDouble1924}, $\mathcal{W}_{p_{\rm init}} = 1/p_{\rm init}$, which is close to the findings of \citet{almeidaTarantulaMassiveBinary2017}.
Unless the index of the period-distribution power-law is positive and high, which is not supported by any observational sample of young, massive stars, short-period binaries are favored over longer ones.
We use no twin binary excess in our setup, since high-mass stars have small twin fractions \citep{sanaBinaryInteractionDominates2012,moeMindYourPs2017}.
The age prior indicates we assume a constant star-formation rate.
\par
\begin{figure*}[h]
	\centering
	\includegraphics[width=0.95\textwidth]{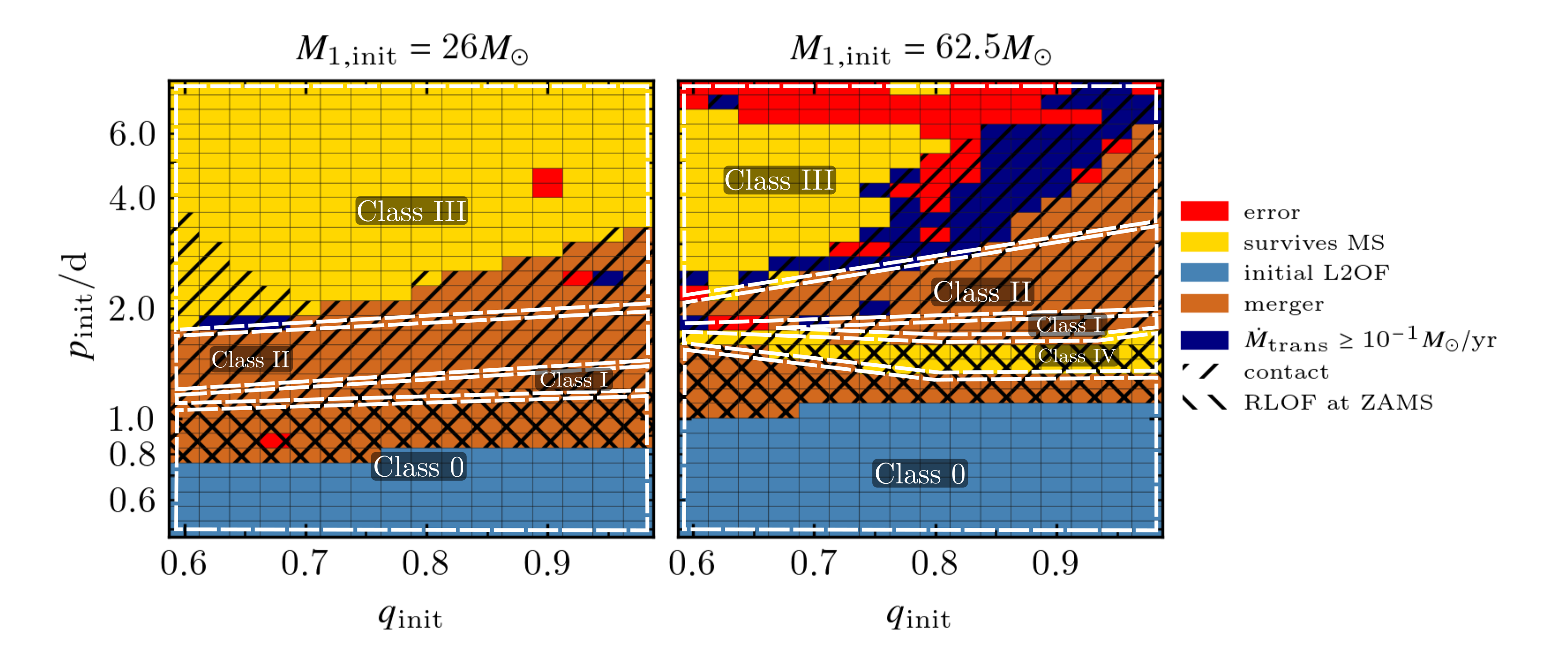}
	\caption{Termination conditions of the $\qty{26}{\Msun}$ and $\qty{62.5}{\Msun}$ models that use ET, as function of the initial mass ratio $q_{\rm init} = M_{2, \rm init}/M_{1, \rm init}$ and initial period $p_{\rm init}$. In color we indicate the models that experience L2OF at the ZAMS (light blue), those that experience L2OF during the main sequence (orange, ``mergers'') or experience a mass transfer rate above $\qty{e-1}{\Msun\per\year}$ (dark blue), those that survive the main sequence without experiencing L2OF (yellow) and those terminate with a numerical error (red). Forward-hatched models experience contact during their main-sequence evolution, while backward-hatched models overflow their RL at the ZAMS. Outlined in white are five classes, ``Class 0-IV'', of systems determined by their qualitative evolution, see Sect.~\ref{ssec:examples}.}
	\label{fig:completion}
\end{figure*}
Given we have a grid of a discrete number of models, each grid point represents a finite area of the birth distributions in Eqs.~\eqref{eqs:birthdists}.
In particular, we compute:
\begin{equation}\label{eq:discrete}
	\int_{M_{\rm lower}}^{M_{\rm upper}} M_{1, \rm init}^{-2.35}\odif{M_{1, \rm init}} \propto M_{1, \rm init, upper}^{-1.35} - M_{1, \rm init, lower}^{-1.35},
\end{equation}
for the contribution of the model of initial primary mass $M_{1, \rm init}$, and the upper and lower bounds are taken in the middle between our grid points.
For points on the edge of the grid, we linearly extrapolate the spacing outward and then use Eq.~\eqref{eq:discrete} as usual.
In this way, the model with initial primary mass $\qty{8}{\Msun}$ has $M_{\rm upper} = \qty{8.5}{\Msun}$ and $M_{\rm lower} = \qty{7.5}{\Msun}$, while the model of highest initial mass has $M_{\rm upper} = \qty{71.125}{\Msun}$ and $M_{\rm lower} = \qty{68.875}{\Msun}$.
Since our grid is logarithmically spaced in initial period, and uniformly in initial mass ratio, given the assumed weights, each model has the same contribution in those dimensions.
\par

With this setup, we can compute the probability of finding a contact binary within a bin of a certain observed period and mass ratio:
\begin{equation}\label{eq:p_q_prob}
	\mathcal{P}_{\rm contact}(q_{\rm obs}, p_{\rm obs}) = \frac{\int \delta(q - q_{\rm obs})\delta(p- p_{\rm obs}) \mathcal{W}_\mathcal{N}\odif{\mathcal{N}}}{\int \mathcal{W}_\mathcal{N}\odif{\mathcal{N}}},
\end{equation}
where we defined
\begin{equation}
	q_{\rm obs} = \min(M_2/M_1, M_1/M_2)
\end{equation}
as the observed mass ratio.
Additionally we filter the models only to times where they are in a contact configuration, $R_{1} \geq R_{\rm RL, 1}$ and $R_{2} \geq R_{\rm RL, 2}$.
\par

 \section{Results and discussion}\label{sec:results}

After computing the binary-evolution models according to the methods described in Sect.~\ref{sec:meth}, we obtain the following outcomes.
Figure \ref{fig:completion} shows the termination conditions of the ET models of initial masses of $\qty{26}{\Msun}$ and $\qty{62.5}{\Msun}$ (the results for the full grids are deferred to Figs.~\ref{fig:et_completion}, \ref{fig:noet_completion} and \ref{fig:single_rot_completion}).
We observe a good completion rate, with less than 9\% of models with ET failing and 5\% of models without ET failing.
Furthermore, the lower-mass models in the range $p_{\rm init} = \qtyrange{1.7}{2.1}{\day}$ that fail do so very close to the TAMS.
Hence we do not expect them to contribute much to the time spent in contact.
\par

We also note that a large fraction of models experience L2OF or RLOF at the ZAMS.
We expect all of these models to interact on the PMS.
Furthermore, for the L2OF systems, we expect these to merge even before hydrogen starts burning.
Those systems would then not be observable as binaries with components on the main sequence, and we classify these as ``Class 0'' in Fig.~\ref{fig:completion}.
Because the models that experience RLOF at the ZAMS were given an ad-hoc treatment to smoothly let MT start as the stars approach to their initial period, we do not include them in the population synthesis of the main text.
We classify them under Class 0 too, and we briefly discuss them in Appendix \ref{app:pms_interaction}.
\par

\subsection{Example models}\label{ssec:examples}
Apart from the models that interact on the PMS, which we called ``Class 0'', we identify four other categories of evolution for our Case A binaries, indicated with ``Class I-IV'' in Fig.~\ref{fig:completion}.
Below, we discuss each of the classes.
\par

\begin{figure}
	\centering
	\includegraphics{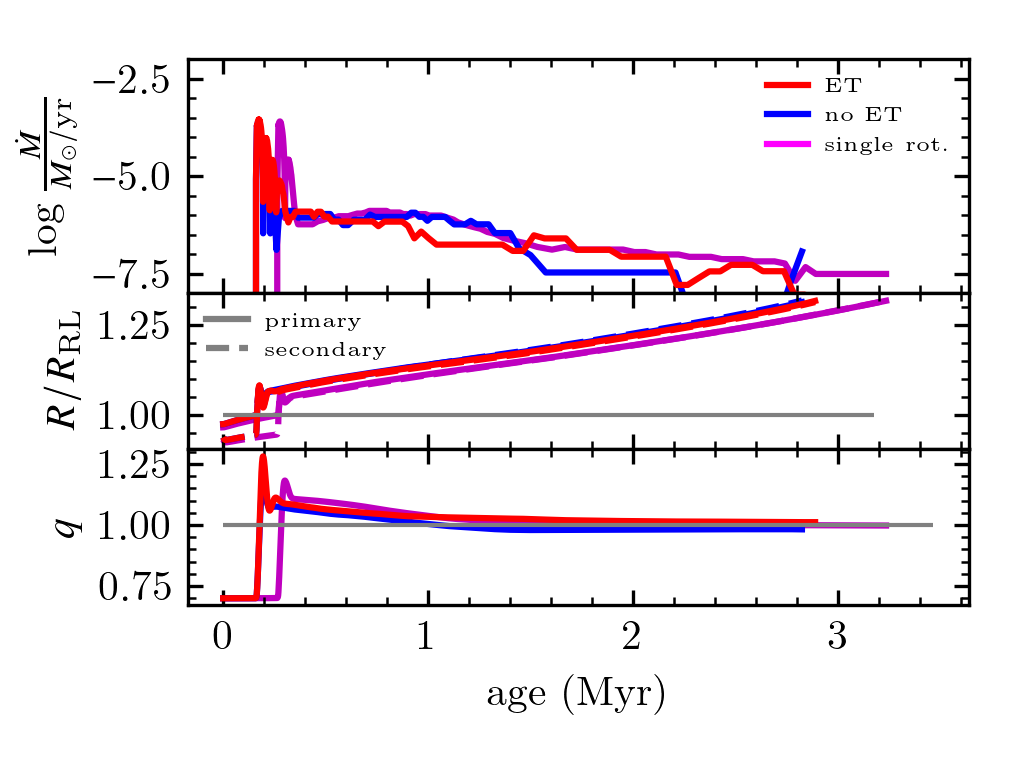}
	\caption{Evolution of a Class I system with initial parameters $M_{1,\rm init} = \qty{24}{\Msun}$, $p_{\rm init} = \qty{1.15}{\day}$ and $q_{\rm init} = 0.7$. The top panel shows the MT rate, the middle panel the relative RL radii, $R/R_{\rm RL}$, and the bottom panel the mass ratio, $q = M_2 / M_1$. This model behaves like System 1 of \citet{menonDetailedEvolutionaryModels2021}, and equalizes masses rapidly irrespective of whether ET is used or not.}
	\label{fig:case1}
\end{figure}

The ``Class I'' models contain those that interact early on the main sequence, which were dubbed ``System 1'' by \citet{menonDetailedEvolutionaryModels2021}.
Here, the more massive star starts fast Case A MT before an appreciable fraction of the main sequence is traversed.
Contact is engaged immediately, and as soon as both stars regain thermal equilibrium after the rapid MT event, MT reverses and the system promptly evolves to mass ratios close to unity.
We show an example of this type of evolution in Fig.~\ref{fig:case1}, which has initial conditions $M_{1,\rm init} = \qty{24}{\Msun}$, $p_{\rm init} = \qty{1.15}{\day}$ and $q_{\rm init} = 0.7$.
We note that ET has little impact on the long-term evolution of this system.
The small age difference seen in the purple single-rotating curve is due to those stars being slightly more compact at the same mass and evolutionary stage \citepalias{fabryModelingOvercontactBinaries2022}, hence interaction occurs slightly later.

\par

\begin{figure}
	\centering
	\includegraphics{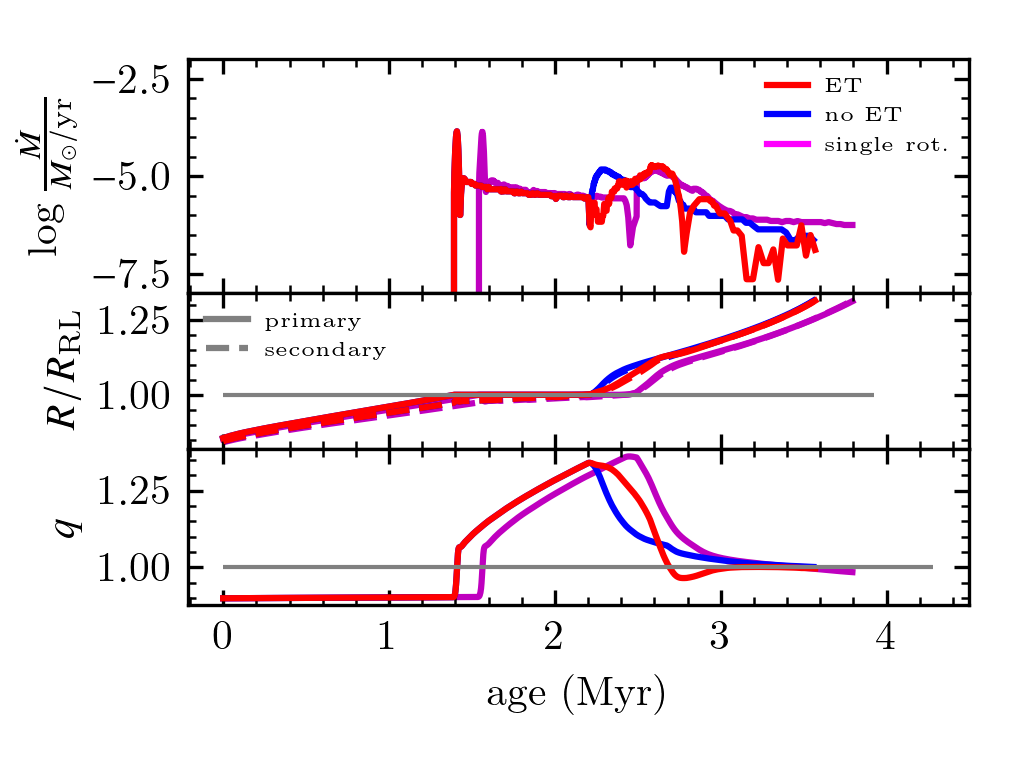}
	\caption{Evolution of a ``Class II'' system with initial parameters $M_{1,\rm init} = \qty{30}{\Msun}$, $p_{\rm init} = \qty{1.52}{\day}$ and $q_{\rm init} = 0.9$, with similar axes as Fig.~\ref{fig:case1}. This system behaves like ``System 2'' of \citet{menonDetailedEvolutionaryModels2021}, and to the case study in \citetalias{fabryModelingContactBinaries2023}.}
	\label{fig:case2}
\end{figure}

The ``Class II'' systems are those of low to moderate mass that start MT after an appreciable fraction of the main sequence has been traversed.
Such systems were dubbed ``System 2'' by \citet{menonDetailedEvolutionaryModels2021}, and we give an example evolution ($M_{\rm 1, init} = \qty{30}{\Msun}$, $q_{\rm init} = 0.9$, $p_{\rm init} = \qty{1.52}{\day}$) in Fig.~\ref{fig:case2}.
Here, fast Case A MT is followed by a slow Case A MT phase from the primary, which drives the mass ratio further off unity.
This is the classical Algol phase, and was studied in detail recently by \citet{senDetailedModelsInteracting2022} \citep[see also, ][]{polsCaseEvolutionMassive1994, vanbeverenWROtypeStar1998, wellsteinFormationContactMassive2001}.
Then, as nuclear-timescale contact is engaged once the accretor fills its RL, MT reverses that eventually equalizes the masses of the components.
This kind of evolution was shown in detail in \citetalias{fabryModelingContactBinaries2023}.
ET has a significant effect for such systems, because it moderates the MT rate at the point where nuclear timescale contact is engaged.
In particular, in both the example given here and in \citetalias{fabryModelingContactBinaries2023}, ET delays the mass equalization, so that, models with ET are more likely to be observed at unequal mass ratios.
Similarly to the Class I example in Fig.~\ref{fig:case1}, the model with the single-rotating star corrections interacts slightly later, but otherwise equalizes its mass ratio as quickly as the no-ET model.

The ``Class III'' systems are those that spend less than 5\% of their total lifetime in contact phases.
They occur at all masses when Case A MT starts late on the main sequence, corresponding to initial orbital periods $p_{\rm init} \gtrsim \qtyrange{2}{3}{\day}$.
Most models in this regime even avoid contact during fast Case A (the exceptions being at low mass and low mass ratio, $M_{1, \rm init} \lesssim \qty{40}{\Msun}$, $q_{\rm init} \lesssim 0.7$).
All models go through a fast Case A MT, and a subsequent Algol phase.
Then, at high initial period, and/or low initial mass ratio, the primary of the system reaches core-hydrogen exhaustion.
Another scenario in this class is that contact is engaged as the accretor in the Algol reaches its RL, but conversely to Class II, the contact phase is limited because the components increase in size significantly as they reach the end of the main sequence.
The models in this category are thus most likely observable as pre-interaction or Algol systems.
\par

\begin{figure}
	\centering
	\includegraphics{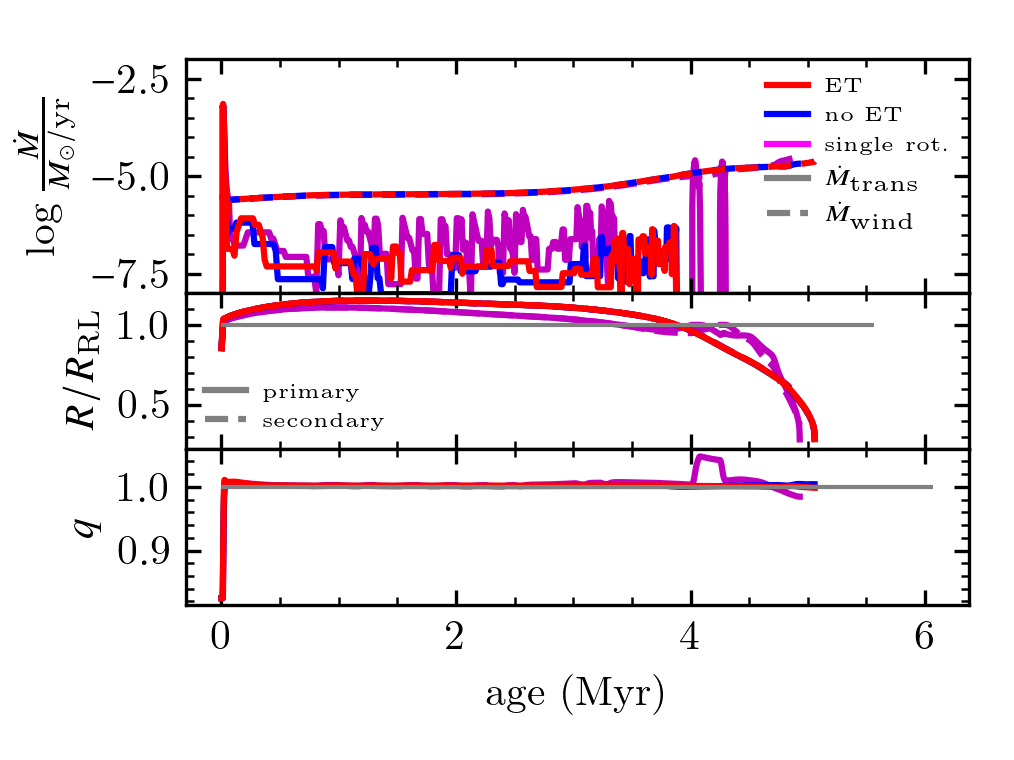}
	\caption{Evolution of a ``Class IV'' system with initial parameters $M_{1,\rm init} = \qty{62.5}{\Msun}$, $p_{\rm init} = \qty{1.52}{\day}$ and $q_{\rm init} = 0.825$. This model undergoes (near-)CHE, and its orbit widens due to significant wind mass loss.}
	\label{fig:case4}
\end{figure}

Finally, the ``Class IV'' systems we distinguish are the (very) massive models ($M_{1, \rm init} \gtrsim \qty{45}{\Msun}$) that interact early on the main sequence.
The models in this regime suffer significant mass and angular momentum loss due to strong stellar winds (recall we model galactic binaries), and they also rotate quickly because of their very short periods and quick synchronization.
Both these properties have the effect of keeping the radius to RL size small, because these models evolve nearly chemically homogeneously, and the wind mass loss causes orbital widening.
An example of this evolution is shown in Fig.~\ref{fig:case4}.
Models in this class engage in nuclear-timescale contact, but eventually the orbit widens enough for the stars to detach, and they avoid a main-sequence merger.
These models are binary black-hole merger progenitors \citep{marchantNewRouteMerging2016}.
For the specific model shown in Fig.~\ref{fig:case4}, the secondary reaches core-hydrogen exhaustion first.

\begin{figure*}
	\centering
	\includegraphics[width=\textwidth]{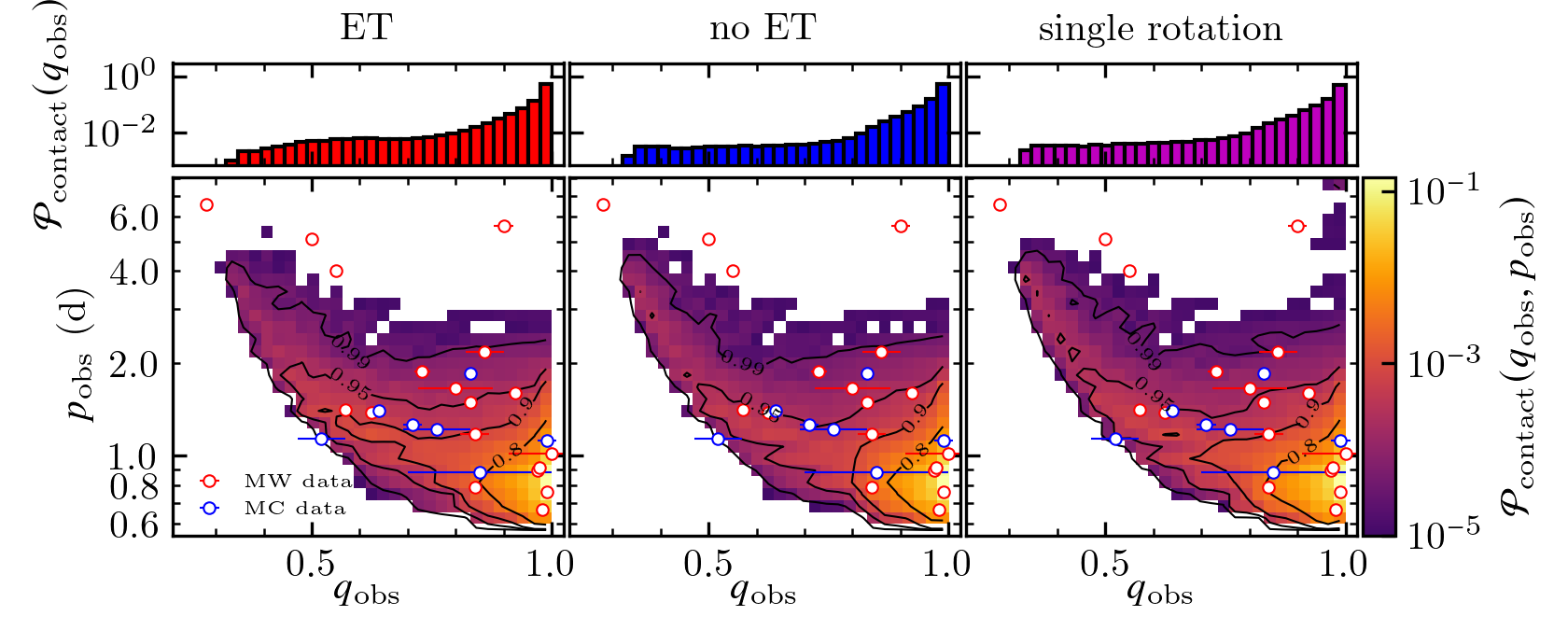}
	\caption{Probability density of massive contact binaries in $(q_{\rm obs}, p_{\rm obs})$ (lower panels) and $q_{\rm obs}$ space (upper panels) for our three model grids.
	The thin black lines are HDIs of the indicated percentiles. The parameters of observed systems from the Milky Way (MW) and Magellanic Clouds (MC) are listed in Table \ref{tab:mw_systems} and \ref{tab:mc_systems}.}
	\label{fig:p_q_dist}
\end{figure*}
\par

\subsection{Population properties}
\begin{figure}
	\centering
	\includegraphics{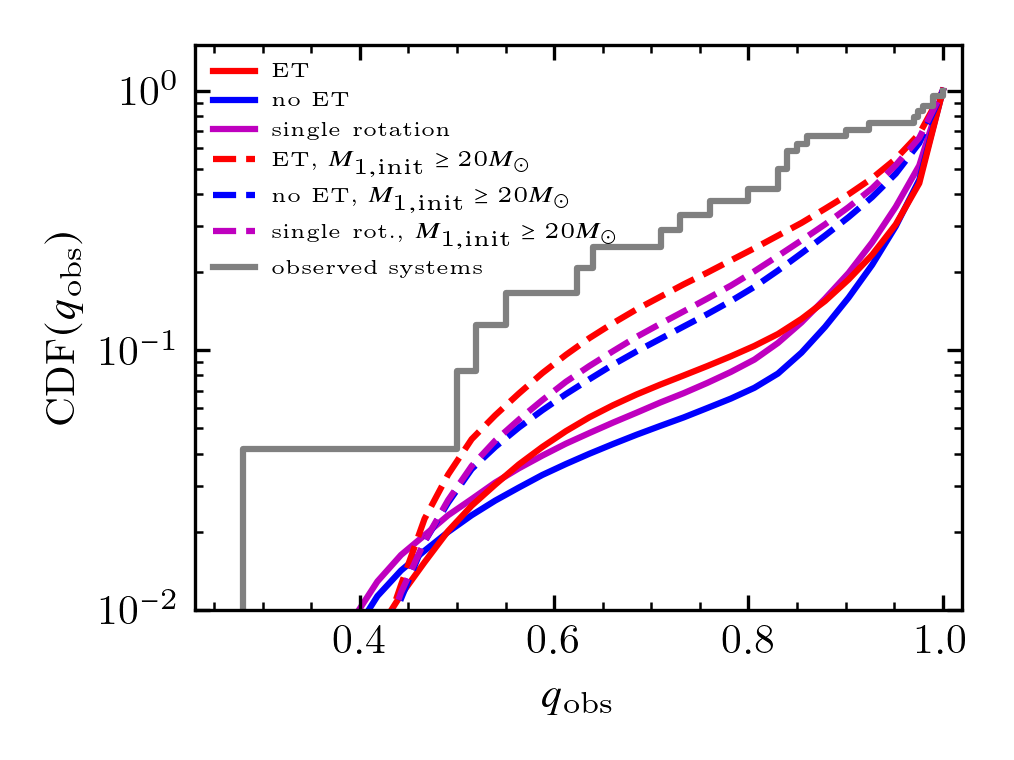}
	\caption{Cumulative distributions as a function of observed mass ratio for the full synthetic populations (full lines), the populations considering only masses above $\qty{20}{\Msun}$ (dashed lines), and the observed sample of Tables \ref{tab:mw_systems} and \ref{tab:mc_systems} (gray line).}
	\label{fig:cdfs}
\end{figure}

From our model outputs, we compute the probability of finding contact systems with a certain observed period and mass ratio.
This amounts to evaluating Eq.~\eqref{eq:p_q_prob} for our grids.
Figure \ref{fig:p_q_dist} shows the result.
First we note that the models in all grids heavily cluster toward the parameters $(q_{\rm obs}, p_{\rm obs}) = (1, \qty{1}{\day})$, similarly to the findings of \citet{menonDetailedEvolutionaryModels2021}.
Overall, the three grids of models have a similar probability density distribution.
To better highlight the differences between the distributions, we have drawn Highest Density Intervals (HDIs) on the panels of Fig.~\ref{fig:p_q_dist} (shown by black lines) to indicate in what regions a certain fraction of models would be found.
Only in the 90th percentile we see a noticeable shift.
Taking the single rotating models as baseline, we see that including the tidal deformation corrections (but no ET) slightly moves the lowest observed mass ratio in toward one, while using tidal reformation and ET moves it toward lower mass ratios.
However, when computing the expected mass ratio, $\bar{q}$, from our PDFs, we get $\bar{q} = 0.929$ for the single-rotating models, $\bar{q} = 0.937$ for the no-ET models, and $\bar{q} = 0.926$ for the ET models.
The effects of tidal deformation and energy transfer are thus very limited when considering the full population.
In Fig.~\ref{fig:cdfs}, we show the cumulative density function (CDF) of the models as function of observed mass ratio, in full lines.
\par

Kolmogorov-Smirnov (KS) tests of the observed sample (Tables \ref{tab:mc_systems} and \ref{tab:mw_systems}) with respect to the ET, no-ET and single-rotation model distributions give $p$-values of $\num{6.2e-7}$, $\num{7.3e-8}$ and $\num{4.8e-7}$ respectively, indicating that it is extremely unlikely these distributions correctly represent the observed sample.
\par

Because the observed systems heavily favor massive O+O contact binaries, in Fig.~\ref{fig:cdfs} we plotted also, in dashed lines, the CDFs when considering only the models with initial primary mass higher than $\qty{20}{\Msun}$.
We note that all grids of models are less clustered toward mass ratios of unity, but despite the shift in the right direction, we still over-predict mass ratios close to unity.
Performing a KS test on the ET models of mass $M_{1, \rm init} > \qty{20}{\Msun}$ returns a $p$-value of 0.003, indicating that, unless we require a confidence higher than 99.7\%, we conclude that the current observed sample is not distributed according to the models.
A KS-test on the no-ET and single-rotating models of high mass gives $p$-values of only \num{1e-4} and \num{5e-4}, respectively.
\par

\begin{figure*}
	\centering
	\includegraphics[width=\textwidth]{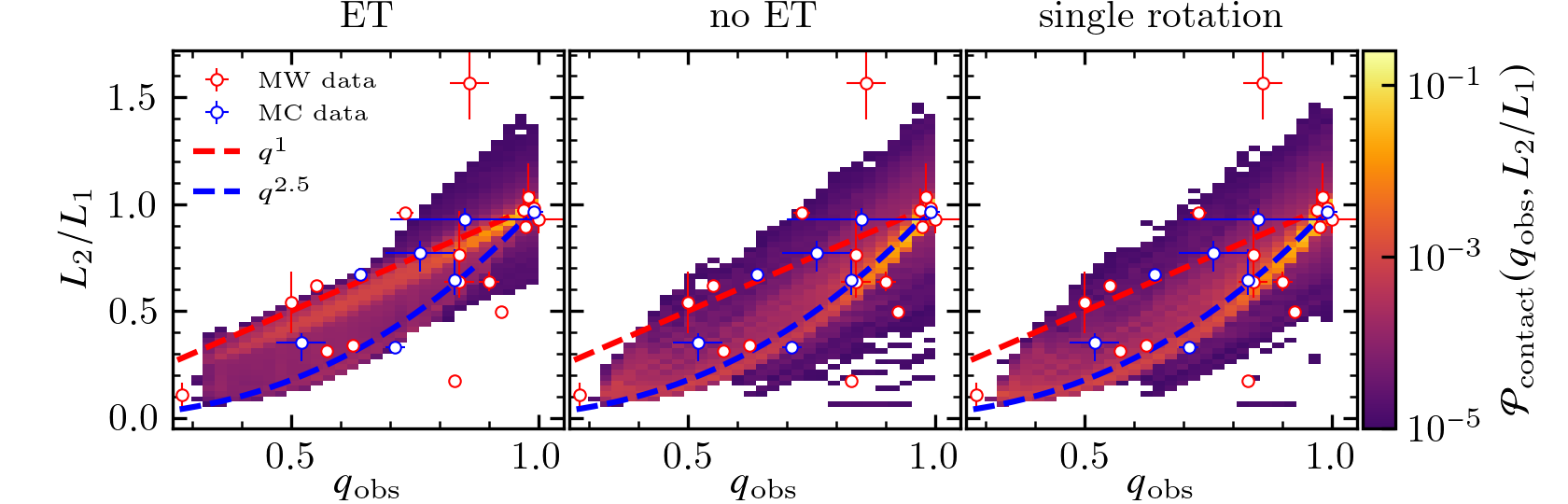}
	\caption{Probability densities in observed mass- and luminosity-ratio space. $L_2/L_1$ here means the ratio of the luminosity of the less massive star to that of the more massive star. We observe a clear difference in the model distributions when ET is applied.}
	\label{fig:lum_ratio_dist}
\end{figure*}
\par

\subsection{Luminosity ratios}\label{ssec:lum_ratio}
Despite not significantly affecting the mass-ratio evolution of massive contact binaries, ET does affect other observable properties of the models.
Since ET raises the luminosity of the energy gainer, and reduces it for the energy donor, we expect a significantly different luminosity ratio of models in contact that include ET versus those that do not.
\par

In Fig.~\ref{fig:lum_ratio_dist}, we plot the probability density as function of the observed mass- and luminosity-ratio of the whole population.
We see a very clear difference between the model distributions that apply ET (left panel) and those that do not (two rightmost panels).
When ET is applied, the models follow closely a $L\sim q$ relation, as expected from the theory discussed in \citetalias{fabryModelingContactBinaries2023}.
The no-ET and single-rotating models instead follow the single-star mass-luminosity relation $L\sim q^{2.5}$.
We refrain from making a formal comparison with the observed sample because it could be contaminated with shallow-contact and near-contact systems.
This is important because, for the luminosity ratio to follow a $L\sim q$ relation, there must be good thermal contact between the components, which might not be the case for shallow contact systems (\citetalias{fabryModelingContactBinaries2023}, Sect.~4.4), and is precluded for (semi-)detached ones.

\subsection{Observational biases in the sample}
The KS tests performed above indicate that, from a statistical point of view, we must reject the hypothesis that the observational sample is distributed per our model distributions.
Still, there could be observational biases that skew the sample.
It could be that the sample of contact binaries is heavily biased toward detecting unequal-mass systems.
However, owing to their short periods and large radial-velocity semi-amplitudes, all systems, regardless of mass ratio, should be picked up in long-term spectroscopic monitoring programs (such as VFTS, \citealp{evansVLTFLAMESTarantulaSurvey2011} and TMBM, \citealp{almeidaTarantulaMassiveBinary2017}).
Photometric characterization is not dependent on mass ratio either.
There thus seem no obvious observational biases in terms of mass ratio in the current sample of massive contact binaries, apart from its completeness.
In terms of primary mass however, characterization of individual systems seems biased toward higher mass.
From Tables \ref{tab:mw_systems} and \ref{tab:mc_systems}, the amount of O+O binaries that were studied far outnumbers the B+B contact systems when compared to a Salpeter IMF.
Comparing the observational sample to models that reach down to initial primary masses of $\qty{8}{\Msun}$ might thus skew the results, since the sample contains an underrepresented number of B+B contact binaries compared to the distribution used in the population modeling.
Still, when limiting our population to those systems with a primary mass higher than $\qty{20}{\Msun}$ makes a small improvement only, as was shown in Fig.~\ref{fig:cdfs}.
Finally, we note that the current sample of observed contact binaries is a very heterogeneous set, where many different observations and analysis tools were used.
This makes it hard to discern the true contact binaries from those that are only near contact.
\citet{menonStudySixtyoneMassive2024} homogeneously studied sixty-one contact binary candidates in the Magellanic Clouds, which, combined with spectroscopic follow-up and advanced analysis methods like in \citet{abdul-masihConstrainingOvercontactPhase2021}, would significantly improve the quality of the sample.
\par

\subsection{Uncertain population distributions}

\begin{figure*}
	\centering
	\includegraphics[width=\textwidth]{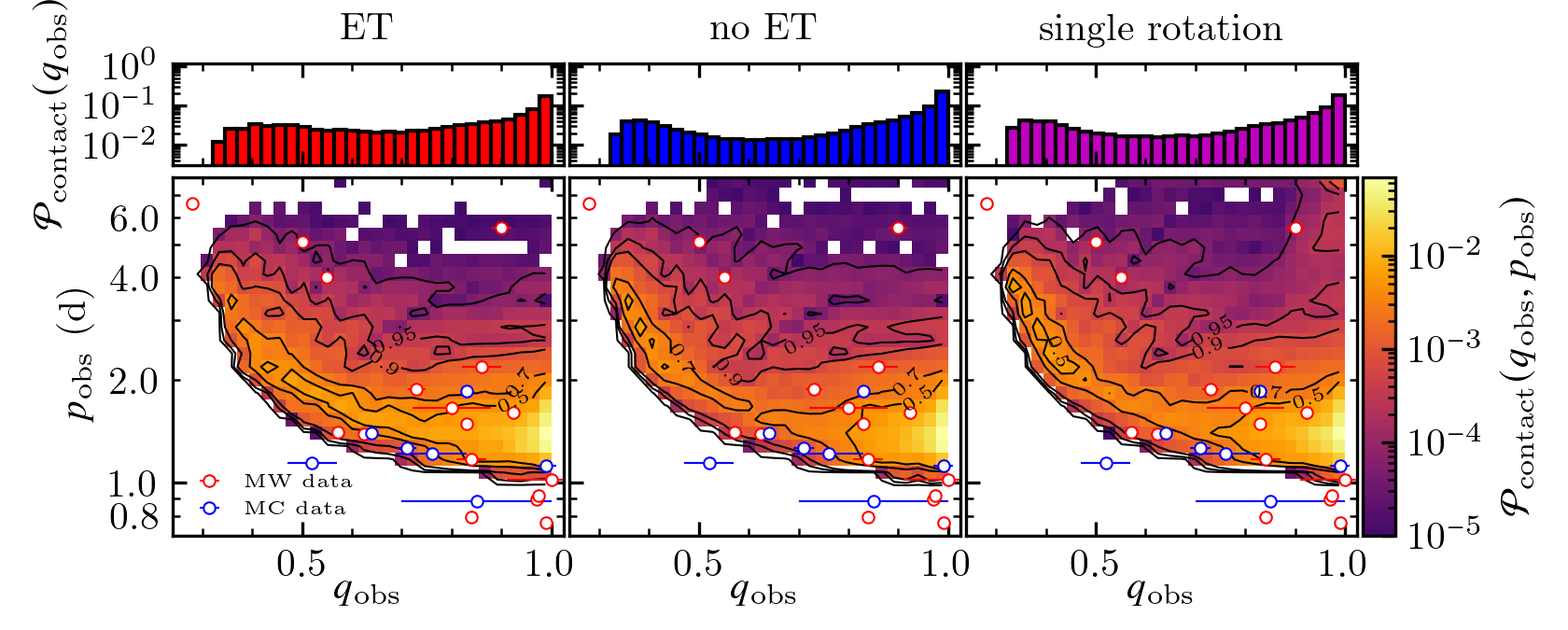}
	\caption{
		Probability density distribution in ($q_{\rm obs}, p_{\rm obs}$) space, for the population with $p_{\rm cut} = \qty{1.26}{\day}$.
		With this period cut, we cannot predict the contact binaries with periods of $p \lesssim \qty{1.2}{\day}$.
	}
	\label{fig:pq_dist_p_cut}
\end{figure*}

A final consideration is the initial period distribution of binaries.
Taking into account the indications that binaries harden over nuclear timescales \citep{sanaDearthShortperiodMassive2017,ramirez-tannusRelationRadialVelocity2021,ramirez-tannusSpectroscopicBinaryFraction2024}, the true ZAMS period distribution potentially does not follow an \"Opik-like law down to periods of one day.
We might even have to consider alternative initial conditions, where one star is at ZAMS, while the other is still on the PMS.
This would require setting new modeling efforts where the star formation process, and the potential interaction with its natal environment on the orbital evolution is included.
\par

With the models computed in this work, we can however test the following.
Like we have done above by considering very high-mass stars alone, here we test whether imposing a low-period cut on the ZAMS period distributions reconciles the models with the observations.
In practice, this means we prevent the systems that interact on the ZAMS (or early on the main sequence) to contribute to the population predictions, effectively restricting the population to Classes II, III and IV only.
\par
\begin{figure}
	\centering
	\includegraphics{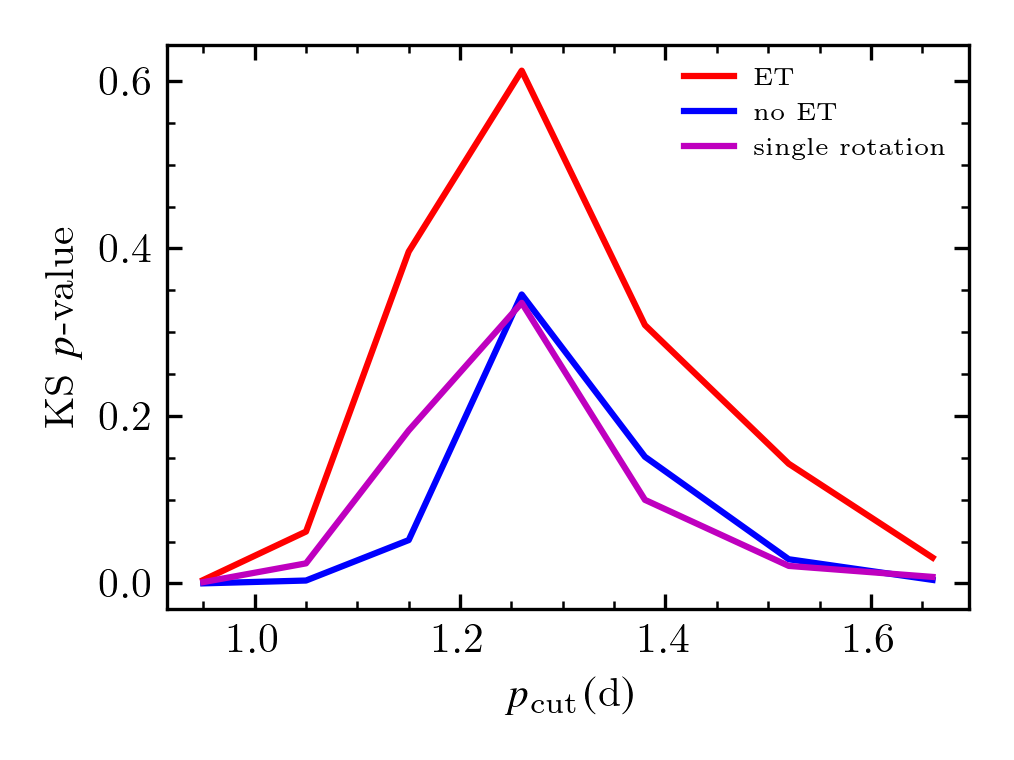}
	\caption{
		KS-test $p$-values when comparing the empirical CDF of the observed sample of massive contact binaries (Tables \ref{tab:mw_systems} and \ref{tab:mc_systems}) against \"Opik initial period distributions with a low-period cut-off $p_{\rm cut}$.
	}
	\label{fig:pcut_pvalues}
\end{figure}

\begin{figure}
\centering
\includegraphics{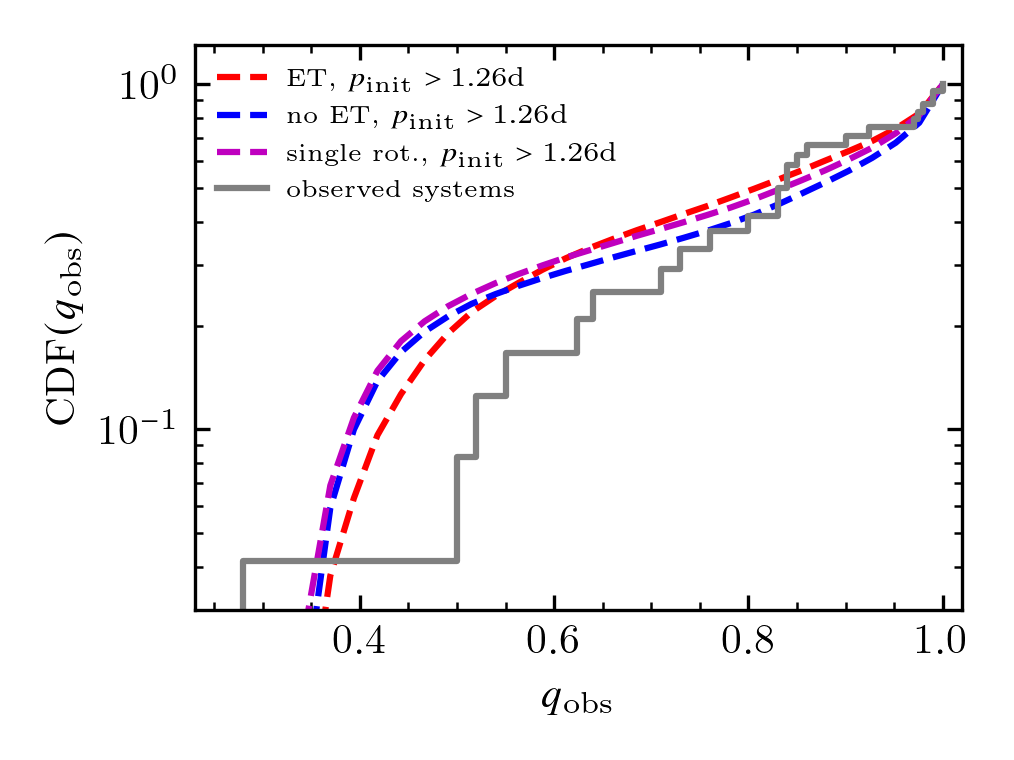}
\caption{
	CDFs of the population as function of observed mass ratio, using the low-period cut-off $p_{\rm cut} = \qty{1.26}{\day}$.
	From the KS statistic (Fig.~\ref{fig:pcut_pvalues}), the observed sample is most likely drawn from the birth distribution with this value for $p_{\rm cut}$.
}
\label{fig:pcut_cumul}
\end{figure}

In Fig.~\ref{fig:pcut_pvalues}, we show the KS-test $p$-values when imposing different short-period cut-offs $p_{\rm cut}$.
We emphasize that this means that the birth distribution probability density is \"Opik's law (Eq.~\ref{eq:p-dist}) down to $p_{\rm cut}$, but zero for $p_{\rm init} \leqslant p_{\rm cut}$.
We see that the population predictions of contact binaries are very sensitive to the assumed birth distribution.
At $p_{\rm cut} = \qty{1.26}{\day}$, the KS statistic of the mass-ratio distribution reaches a minimum of 0.15 for the ET-models, while it is 0.18 for both the no-ET and single-rotating models.
This corresponds to $p$-values of 0.53 and 0.30, respectively, indicating that, to high confidence, we cannot reject the hypothesis that the observed sample is drawn from the distribution with this period cut.
Figure \ref{fig:pcut_cumul} shows the corresponding cumulative probability distributions, and points out that the models now overestimate the extreme-mass-ratio contact binaries.
We note, however, that by placing this cut we cannot explain the period distribution of the sample, see Fig.~\ref{fig:pq_dist_p_cut}.
This result suggests that systems with periods of $p \lesssim p_{\rm cut}$ at ZAMS might not occur in nature, and instead might have merged before we observe them.
To make definite statements however would require an in-depth study of the impact on including alternate ZAMS conditions to a population, and should include potential interactions with the star-formation environment.
\par

 \section{Conclusions}\label{sec:conc}
In this work, we have computed Case A interacting binary models and constructed synthetic populations of short-period, massive binary systems.
We find that the inclusion of the effect of efficient ET in the shared envelopes of contact binaries does not fully reconcile the model predictions with the observed sample, in particular on their mass-ratio distribution.
Many of the models avoid nuclear-timescale contact phases (those of Class III in Sect.~\ref{ssec:examples}), or enter contact so early on the main sequence that they equalize in mass within a thermal timescale (Class I), regardless of whether ET is applied or not.
ET has a significant effect only on a small fraction of the models (Class II), where the equalization is delayed as ET modifies the internal structure of the components.
The energy gainer bloats while the energy donor shrinks, slightly altering the mass-radius relationship, which impacts whether a thermally stable contact configuration is possible.
However, the combination of nuclear evolution and efficient rejuvenation of the stellar cores due to overshooting results in the mass ratio evolving to values of unity reasonably quickly still.
\par

From these points we conclude that unequal-mass contact binaries must be reasonably evolved systems.
More specifically, at least one of the components must be significantly departed from the ZAMS, so that its mass-radius relationship is closer to that of the Roche-lobe geometry.
Rejuvenation of the accretor during mass transfer events might not be as efficient as modeled here \citep{braunEffectsAccretionMassive1995}, and this likely impacts the results.
In work by \citet{forsterEvolutionMassiveContact2023}, it is shown that changing the near-core mixing properties, such as the overshooting strength, semi-convective efficiency and convective penetration of molecular weight barriers, can have a large effect on the evolution of contact binaries.
Near-core mixing properties of massive stars are uncertain, but asteroseismic analysis can probe the near-core region, and constrain, for example, overshooting parameters \citep[e.g.,][]{pedersenInternalMixingRotating2021, michielsenProbingTemperatureGradient2021, burssensCalibrationPointStellar2023, johnstonModellingTimedependentConvective2024, bellingerPotentialAsteroseismologyResolve2024}.
Recently, it has been identified that stellar mergers, which have different internal mixing properties than genuine single stars, should reveal clear asteroseismic signatures \citep{hennecoMergerSeismologyDistinguishing2024}.
\par

We find indications that contact-binary progenitors could be affected by orbital hardening over their lifetimes.
First, our models show that the very short-period binaries that interact early on the main sequence, when the stars are still relatively homogeneous, do not form contact binaries of unequal mass.
Since we observe a fair number of such unequal systems, this suggests the closest binaries, which predominantly form equal-mass contact binaries, merge before being observed or even before they reach the ZAMS (like the Class 0 models).
Secondly, when we restrict our population to models of Class II and III only, we better fit the observed mass-ratio distribution.
This suggests that the initial-period population could be distributed to somewhat longer periods, allowing time for the components to evolve from the ZAMS, and shrink on the nuclear timescale to become contact binaries we observe today.
Both these points indicate that the initial period distribution (at ZAMS) is likely not \"Opik's law down to periods shorter than $p\lesssim \qty{2}{\day}$.
Further models of massive binaries, which allow for interaction on the PMS and/or include a prescription for orbital perturbations by the natal environment, are necessary to study this.
\par

Finally, we advocate for renewed efforts to characterize massive contact binaries.
Not only is the sample size rather limited, it is also heterogeneous.
In order to come to a comprehensive sample of confirmed contact binaries, studies like that of \citet{menonStudySixtyoneMassive2024} should be undertaken and followed-up with homogeneous spectroscopic analyses like \citet{abdul-masihConstrainingOvercontactPhase2021} to clearly distinguish the near-contact binaries from the true contact ones.
The modeling of massive contact binaries, from this work or in \citet{menonDetailedEvolutionaryModels2021, menonStudySixtyoneMassive2024} has reached a good level, and we are currently limited by observational constraints on contact binaries to further unravel their evolution.
\par

\begin{acknowledgements}
M.F.~thanks the Flemish research foundation (FWO, Fonds voor Wetenschappelijk Onderzoek) PhD fellowship No.~11H2421N for its support.
P.M.~acknowledges support from the FWO senior postdoctoral fellowship No.~12ZY523N.
This work has made use of the computing resources of the Vlaams Supercomputer Center (VSC), funded by the FWO.
The research leading to these results has received funding from the European Research Council (ERC) under the European Union's Horizon 2020 research and innovation program (grant agreement numbers 772225: MULTIPLES).
Data processing and visualization has been made possible thanks to \texttt{matplotlib} \citep{hunterMatplotlib2DGraphics2007}, \texttt{scipy} \citep{virtanenSciPyFundamentalAlgorithms2020} and \texttt{numpy} \citep{harrisArrayProgrammingNumPy2020}.
\end{acknowledgements}

\bibliographystyle{aa}
\bibliography{massive_stars.bib}

\begin{thebibliography}{150}
\expandafter\ifx\csname natexlab\endcsname\relax\def\natexlab#1{#1}\fi

\bibitem[{{Abdul-Masih} {et~al.}(2022){Abdul-Masih}, Escorza, Menon, Mahy, \&
  Marchant}]{abdul-masihConstrainingOvercontactPhase2022}
{Abdul-Masih}, M., Escorza, A., Menon, A., Mahy, L., \& Marchant, P. 2022,
  Astronomy \& Astrophysics, 666, A18 \csname
  abdul-masihConstrainingOvercontactPhase2022link\endcsname~\csname
  abdul-masihConstrainingOvercontactPhase2022note\endcsname

\bibitem[{{Abdul-Masih} {et~al.}(2020){Abdul-Masih}, Sana, Conroy, Sundqvist,
  Pr{\v s}a, Kochoska, \& Puls}]{abdul-masihSpectroscopicPatchModel2020}
{Abdul-Masih}, M., Sana, H., Conroy, K.~E., {et~al.} 2020, Astronomy \&
  Astrophysics, 636, A59 \csname
  abdul-masihSpectroscopicPatchModel2020link\endcsname~\csname
  abdul-masihSpectroscopicPatchModel2020note\endcsname

\bibitem[{{Abdul-Masih} {et~al.}(2021){Abdul-Masih}, Sana, Hawcroft, Almeida,
  Brands, {de Mink}, Justham, Langer, Mahy, Marchant, Menon, Puls, \&
  Sundqvist}]{abdul-masihConstrainingOvercontactPhase2021}
{Abdul-Masih}, M., Sana, H., Hawcroft, C., {et~al.} 2021, Astronomy \&
  Astrophysics, 651, A96 \csname
  abdul-masihConstrainingOvercontactPhase2021link\endcsname~\csname
  abdul-masihConstrainingOvercontactPhase2021note\endcsname

\bibitem[{{Aguilera-Dena} {et~al.}(2020){Aguilera-Dena}, Langer, Antoniadis, \&
  M{\"u}ller}]{aguilera-denaPrecollapsePropertiesSuperluminous2020}
{Aguilera-Dena}, D.~R., Langer, N., Antoniadis, J., \& M{\"u}ller, B. 2020, The
  Astrophysical Journal, 901, 114 \csname
  aguilera-denaPrecollapsePropertiesSuperluminous2020link\endcsname~\csname
  aguilera-denaPrecollapsePropertiesSuperluminous2020note\endcsname

\bibitem[{{Aguilera-Dena} {et~al.}(2018){Aguilera-Dena}, Langer, Moriya, \&
  Schootemeijer}]{aguilera-denaRelatedProgenitorModels2018}
{Aguilera-Dena}, D.~R., Langer, N., Moriya, T.~J., \& Schootemeijer, A. 2018,
  The Astrophysical Journal, 858, 115 \csname
  aguilera-denaRelatedProgenitorModels2018link\endcsname~\csname
  aguilera-denaRelatedProgenitorModels2018note\endcsname

\bibitem[{Alcock {et~al.}(1997)Alcock, Allsman, Alves, Axelrod, Becker,
  Bennett, Cook, Freeman, Griest, Lacy, Lehner, Marshall, Minniti, Peterson,
  Pratt, Quinn, Rodgers, Stubbs, Sutherland, \&
  Welch}]{alcockMACHOProjectLMC1997}
Alcock, C., Allsman, R.~A., Alves, D., {et~al.} 1997, The Astronomical Journal,
  114, 326 \csname alcockMACHOProjectLMC1997link\endcsname~\csname
  alcockMACHOProjectLMC1997note\endcsname

\bibitem[{Almeida {et~al.}(2015)Almeida, Sana, {de Mink}, Tramper,
  Soszy{\'n}ski, Langer, Barb{\'a}, Cantiello, Damineli, {de Koter}, Garcia,
  Gr{\"a}fener, Herrero, Howarth, Ma{\'i}z~Apell{\'a}niz, Norman,
  {Ram{\'i}rez-Agudelo}, \& Vink}]{almeidaDiscoveryMassiveOvercontact2015}
Almeida, L.~A., Sana, H., {de Mink}, S.~E., {et~al.} 2015, The Astrophysical
  Journal, 812, 102 \csname
  almeidaDiscoveryMassiveOvercontact2015link\endcsname~\csname
  almeidaDiscoveryMassiveOvercontact2015note\endcsname

\bibitem[{Almeida {et~al.}(2017)Almeida, Sana, Taylor, Barb{\'a}, Bonanos,
  Crowther, Damineli, {de Koter}, {de Mink}, Evans, Gieles, Grin,
  {H{\'e}nault-Brunet}, Langer, Lennon, Lockwood, Ma{\'i}z~Apell{\'a}niz,
  Moffat, Neijssel, Norman, {Ram{\'i}rez-Agudelo}, Richardson, Schootemeijer,
  Shenar, Soszy{\'n}ski, Tramper, \& Vink}]{almeidaTarantulaMassiveBinary2017}
Almeida, L.~A., Sana, H., Taylor, W., {et~al.} 2017, Astronomy \& Astrophysics,
  598, A84 \csname almeidaTarantulaMassiveBinary2017link\endcsname~\csname
  almeidaTarantulaMassiveBinary2017note\endcsname

\bibitem[{Angulo {et~al.}(1999)Angulo, Arnould, Rayet, Descouvemont, Baye,
  {Leclercq-Willain}, Coc, Barhoumi, Aguer, Rolfs, Kunz, Hammer, Mayer,
  Paradellis, Kossionides, Chronidou, Spyrou, {degl'Innocenti}, Fiorentini,
  Ricci, Zavatarelli, Providencia, Wolters, Soares, Grama, Rahighi, Shotter, \&
  Lamehi~Rachti}]{anguloCompilationChargedparticleInduced1999}
Angulo, C., Arnould, M., Rayet, M., {et~al.} 1999, Nuclear Physics A, 656, 3
  \csname anguloCompilationChargedparticleInduced1999link\endcsname~\csname
  anguloCompilationChargedparticleInduced1999note\endcsname

\bibitem[{Asplund {et~al.}(2009)Asplund, Grevesse, Sauval, \&
  Scott}]{asplundChemicalCompositionSun2009}
Asplund, M., Grevesse, N., Sauval, A.~J., \& Scott, P. 2009, Annual Review of
  Astronomy and Astrophysics, 47, 481 \csname
  asplundChemicalCompositionSun2009link\endcsname~\csname
  asplundChemicalCompositionSun2009note\endcsname

\bibitem[{Banyard {et~al.}(2022)Banyard, Sana, Mahy, Bodensteiner,
  Villase{\~n}or, \& Evans}]{banyardObservedMultiplicityProperties2022}
Banyard, G., Sana, H., Mahy, L., {et~al.} 2022, Astronomy \& Astrophysics, 658,
  A69 \csname banyardObservedMultiplicityProperties2022link\endcsname~\csname
  banyardObservedMultiplicityProperties2022note\endcsname

\bibitem[{Bastian {et~al.}(2010)Bastian, Covey, \&
  Meyer}]{bastianUniversalStellarInitial2010}
Bastian, N., Covey, K.~R., \& Meyer, M.~R. 2010, Annual Review of Astronomy and
  Astrophysics, 48, 339 \csname
  bastianUniversalStellarInitial2010link\endcsname~\csname
  bastianUniversalStellarInitial2010note\endcsname

\bibitem[{Bellinger {et~al.}(2024)Bellinger, {de Mink}, {van Rossem}, \&
  Justham}]{bellingerPotentialAsteroseismologyResolve2024}
Bellinger, E.~P., {de Mink}, S.~E., {van Rossem}, W.~E., \& Justham, S. 2024,
  The Astrophysical Journal, 967, L39 \csname
  bellingerPotentialAsteroseismologyResolve2024link\endcsname~\csname
  bellingerPotentialAsteroseismologyResolve2024note\endcsname

\bibitem[{Biermann \& Thomas(1972)}]{biermannModelsContactBinaries1972}
Biermann, P. \& Thomas, H.~C. 1972, Astronomy \& Astrophysics, 16, 60 \csname
  biermannModelsContactBinaries1972link\endcsname~\csname
  biermannModelsContactBinaries1972note\endcsname

\bibitem[{Binnendijk(1970)}]{binnendijkOrbitalElementsUrsae1970}
Binnendijk, L. 1970, Vistas in Astronomy, 12, 217 \csname
  binnendijkOrbitalElementsUrsae1970link\endcsname~\csname
  binnendijkOrbitalElementsUrsae1970note\endcsname

\bibitem[{Blouin {et~al.}(2020)Blouin, Shaffer, Saumon, \&
  Starrett}]{blouinNewConductiveOpacities2020}
Blouin, S., Shaffer, N.~R., Saumon, D., \& Starrett, C.~E. 2020, The
  Astrophysical Journal, 899, 46 \csname
  blouinNewConductiveOpacities2020link\endcsname~\csname
  blouinNewConductiveOpacities2020note\endcsname

\bibitem[{Bodensteiner {et~al.}(2022)Bodensteiner, Heida, {Abdul-Masih}, Baade,
  Banyard, Bowman, Fabry, Frost, Mahy, Marchant, M{\'e}rand, Reggiani,
  Rivinius, Sana, Selman, \& Shenar}]{bodensteinerDetectingStrippedStars2022}
Bodensteiner, J., Heida, M., {Abdul-Masih}, M., {et~al.} 2022, The Messenger,
  186, 3 \csname bodensteinerDetectingStrippedStars2022link\endcsname~\csname
  bodensteinerDetectingStrippedStars2022note\endcsname

\bibitem[{{B{\"o}hm-Vitense}(1958)}]{bohm-vitenseUberWasserstoffkonvektionszoneSternen1958}
{B{\"o}hm-Vitense}, E. 1958, Zeitschrift fur Astrophysik, 46, 108 \csname
  bohm-vitenseUberWasserstoffkonvektionszoneSternen1958link\endcsname~\csname
  bohm-vitenseUberWasserstoffkonvektionszoneSternen1958note\endcsname

\bibitem[{Bondi \& Hoyle(1944)}]{bondiMechanismAccretionStars1944}
Bondi, H. \& Hoyle, F. 1944, Monthly Notices of the Royal Astronomical Society,
  104, 273 \csname bondiMechanismAccretionStars1944link\endcsname~\csname
  bondiMechanismAccretionStars1944note\endcsname

\bibitem[{Braun \& Langer(1995)}]{braunEffectsAccretionMassive1995}
Braun, H. \& Langer, N. 1995, Astronomy and Astrophysics, 297, 483 \csname
  braunEffectsAccretionMassive1995link\endcsname~\csname
  braunEffectsAccretionMassive1995note\endcsname

\bibitem[{Brott {et~al.}(2011)Brott, de~Mink, Cantiello, Langer, de~Koter,
  Evans, Hunter, Trundle, \& Vink}]{brottRotatingMassiveMainsequence2011}
Brott, I., de~Mink, S.~E., Cantiello, M., {et~al.} 2011, Astronomy \&
  Astrophysics, 530, A115 \csname
  brottRotatingMassiveMainsequence2011link\endcsname~\csname
  brottRotatingMassiveMainsequence2011note\endcsname

\bibitem[{Burssens {et~al.}(2023)Burssens, Bowman, Michielsen,
  {Sim{\'o}n-D{\'i}az}, Aerts, Vanlaer, Banyard, Nardetto, Townsend, Handler,
  Mombarg, Vanderspek, \& Ricker}]{burssensCalibrationPointStellar2023}
Burssens, S., Bowman, D.~M., Michielsen, M., {et~al.} 2023, Nature Astronomy,
  7, 913 \csname burssensCalibrationPointStellar2023link\endcsname~\csname
  burssensCalibrationPointStellar2023note\endcsname

\bibitem[{{\c C}ak{\i}rl{\i} {et~al.}(2014){\c C}ak{\i}rl{\i}, Ibanoglu, \&
  Sipahi}]{cakirliV745CassiopeanInteracting2014}
{\c C}ak{\i}rl{\i}, {\"O}., Ibanoglu, C., \& Sipahi, E. 2014, Monthly Notices
  of the Royal Astronomical Society, 442, 1560 \csname
  cakirliV745CassiopeanInteracting2014link\endcsname~\csname
  cakirliV745CassiopeanInteracting2014note\endcsname

\bibitem[{Cassisi {et~al.}(2007)Cassisi, Potekhin, Pietrinferni, Catelan, \&
  Salaris}]{cassisiUpdatedElectronconductionOpacities2007}
Cassisi, S., Potekhin, A.~Y., Pietrinferni, A., Catelan, M., \& Salaris, M.
  2007, The Astrophysical Journal, 661, 1094 \csname
  cassisiUpdatedElectronconductionOpacities2007link\endcsname~\csname
  cassisiUpdatedElectronconductionOpacities2007note\endcsname

\bibitem[{Chugunov {et~al.}(2007)Chugunov, Dewitt, \&
  Yakovlev}]{chugunovCoulombTunnelingFusion2007}
Chugunov, A.~I., Dewitt, H.~E., \& Yakovlev, D.~G. 2007, Physical Review D, 76,
  025028 \csname chugunovCoulombTunnelingFusion2007link\endcsname~\csname
  chugunovCoulombTunnelingFusion2007note\endcsname

\bibitem[{Cox \& Giuli(1968)}]{coxPrinciplesStellarStructure1968}
Cox, J.~P. \& Giuli, R.~T. 1968, Principles of Stellar Structure (New York:
  {Gordon and Breach}) \csname
  coxPrinciplesStellarStructure1968link\endcsname~\csname
  coxPrinciplesStellarStructure1968note\endcsname

\bibitem[{Cyburt {et~al.}(2010)Cyburt, Amthor, Ferguson, Meisel, Smith, Warren,
  Heger, Hoffman, Rauscher, Sakharuk, Schatz, Thielemann, \&
  Wiescher}]{cyburtJINAREACLIBDatabase2010}
Cyburt, R.~H., Amthor, A.~M., Ferguson, R., {et~al.} 2010, The Astrophysical
  Journal Supplement Series, 189, 240 \csname
  cyburtJINAREACLIBDatabase2010link\endcsname~\csname
  cyburtJINAREACLIBDatabase2010note\endcsname

\bibitem[{{de Mink} {et~al.}(2013){de Mink}, Langer, Izzard, Sana, \& {de
  Koter}}]{deminkRotationRatesMassive2013}
{de Mink}, S.~E., Langer, N., Izzard, R.~G., Sana, H., \& {de Koter}, A. 2013,
  The Astrophysical Journal, 764, 166 \csname
  deminkRotationRatesMassive2013link\endcsname~\csname
  deminkRotationRatesMassive2013note\endcsname

\bibitem[{Drout {et~al.}(2023)Drout, G{\"o}tberg, Ludwig, Groh, {de Mink},
  O'Grady, \& Smith}]{droutObservedPopulationIntermediatemass2023}
Drout, M.~R., G{\"o}tberg, Y., Ludwig, B.~A., {et~al.} 2023, Science, 382, 1287
  \csname droutObservedPopulationIntermediatemass2023link\endcsname~\csname
  droutObservedPopulationIntermediatemass2023note\endcsname

\bibitem[{Dunstall {et~al.}(2015)Dunstall, Dufton, Sana, Evans, Howarth,
  {Sim{\'o}n-D{\'i}az}, {de Mink}, Langer, Ma{\'i}z~Apell{\'a}niz, \&
  Taylor}]{dunstallVLTFLAMESTarantulaSurvey2015}
Dunstall, P.~R., Dufton, P.~L., Sana, H., {et~al.} 2015, Astronomy \&
  Astrophysics, 580, A93 \csname
  dunstallVLTFLAMESTarantulaSurvey2015link\endcsname~\csname
  dunstallVLTFLAMESTarantulaSurvey2015note\endcsname

\bibitem[{Eddington(1925)}]{eddingtonCirculatingCurrentsRotating1925}
Eddington, A.~S. 1925, The Observatory, 48, 73 \csname
  eddingtonCirculatingCurrentsRotating1925link\endcsname~\csname
  eddingtonCirculatingCurrentsRotating1925note\endcsname

\bibitem[{Eggen(1967)}]{eggenContactBinariesII1967}
Eggen, O.~J. 1967, Memoirs of the Royal Astronomical Society, 70, 111 \csname
  eggenContactBinariesII1967link\endcsname~\csname
  eggenContactBinariesII1967note\endcsname

\bibitem[{Eggleton(1983)}]{eggletonApproximationsRadiiRoche1983}
Eggleton, P.~P. 1983, The Astrophysical Journal, 268, 368 \csname
  eggletonApproximationsRadiiRoche1983link\endcsname~\csname
  eggletonApproximationsRadiiRoche1983note\endcsname

\bibitem[{Endal \& Sofia(1978)}]{endalEvolutionRotatingStars1978}
Endal, A.~S. \& Sofia, S. 1978, The Astrophysical Journal, 220, 279 \csname
  endalEvolutionRotatingStars1978link\endcsname~\csname
  endalEvolutionRotatingStars1978note\endcsname

\bibitem[{Evans {et~al.}(2011)Evans, Taylor, {H{\'e}nault-Brunet}, Sana, {de
  Koter}, {Sim{\'o}n-D{\'i}az}, Carraro, Bagnoli, Bastian, Bestenlehner,
  Bonanos, Bressert, Brott, Campbell, Cantiello, Clark, Costa, Crowther, {de
  Mink}, Doran, Dufton, Dunstall, Friedrich, Garcia, Gieles, Gr{\"a}fener,
  Herrero, Howarth, Izzard, Langer, Lennon, Ma{\'i}z~Apell{\'a}niz, Markova,
  Najarro, Puls, Ramirez, {Sab{\'i}n-Sanjuli{\'a}n}, Smartt, Stroud, {van
  Loon}, Vink, \& Walborn}]{evansVLTFLAMESTarantulaSurvey2011}
Evans, C.~J., Taylor, W.~D., {H{\'e}nault-Brunet}, V., {et~al.} 2011, Astronomy
  \& Astrophysics, 530, A108 \csname
  evansVLTFLAMESTarantulaSurvey2011link\endcsname~\csname
  evansVLTFLAMESTarantulaSurvey2011note\endcsname

\bibitem[{Fabry {et~al.}(2023)Fabry, Marchant, Langer, \&
  Sana}]{fabryModelingContactBinaries2023}
Fabry, M., Marchant, P., Langer, N., \& Sana, H. 2023, Astronomy \&
  Astrophysics, 672, A175 \csname
  fabryModelingContactBinaries2023link\endcsname~\csname
  fabryModelingContactBinaries2023note\endcsname

\bibitem[{Fabry {et~al.}(2022)Fabry, Marchant, \&
  Sana}]{fabryModelingOvercontactBinaries2022}
Fabry, M., Marchant, P., \& Sana, H. 2022, Astronomy \& Astrophysics, 661, A123
  \csname fabryModelingOvercontactBinaries2022link\endcsname~\csname
  fabryModelingOvercontactBinaries2022note\endcsname

\bibitem[{Ferguson {et~al.}(2005)Ferguson, Alexander, Allard, Barman, Bodnarik,
  Hauschildt, {Heffner-Wong}, \& Tamanai}]{fergusonLowTemperatureOpacities2005}
Ferguson, J.~W., Alexander, D.~R., Allard, F., {et~al.} 2005, The Astrophysical
  Journal, 623, 585 \csname
  fergusonLowTemperatureOpacities2005link\endcsname~\csname
  fergusonLowTemperatureOpacities2005note\endcsname

\bibitem[{Flannery(1976)}]{flanneryCyclicThermalInstability1976}
Flannery, B.~P. 1976, The Astrophysical Journal, 205, 217 \csname
  flanneryCyclicThermalInstability1976link\endcsname~\csname
  flanneryCyclicThermalInstability1976note\endcsname

\bibitem[{F{\"o}rster(2023)}]{forsterEvolutionMassiveContact2023}
F{\"o}rster, K.~U. 2023, Evolution of {{Massive Contact Binaries}},
  https://astro.uni-bonn.de/{\textasciitilde}nlanger/thesis/kai\_bachelor.pdf
  \csname forsterEvolutionMassiveContact2023link\endcsname~\csname
  forsterEvolutionMassiveContact2023note\endcsname

\bibitem[{Fricke(1968)}]{frickeInstabilitatStationarerRotation1968}
Fricke, K. 1968, Zeitschrift fur Astrophysik, 68, 317 \csname
  frickeInstabilitatStationarerRotation1968link\endcsname~\csname
  frickeInstabilitatStationarerRotation1968note\endcsname

\bibitem[{Frost {et~al.}(2024)Frost, Sana, Mahy, Wade, Barron, Le~Bouquin,
  M{\'e}rand, Schneider, Shenar, Barb{\'a}, Bowman, Fabry, Farhang, Marchant,
  Morrell, \& Smoker}]{frostMagneticMassiveStar2024}
Frost, A.~J., Sana, H., Mahy, L., {et~al.} 2024, Science, 384, 214 \csname
  frostMagneticMassiveStar2024link\endcsname~\csname
  frostMagneticMassiveStar2024note\endcsname

\bibitem[{Fuller {et~al.}(1985)Fuller, Fowler, \&
  Newman}]{fullerStellarWeakInteraction1985}
Fuller, G.~M., Fowler, W.~A., \& Newman, M.~J. 1985, The Astrophysical Journal,
  293, 1 \csname fullerStellarWeakInteraction1985link\endcsname~\csname
  fullerStellarWeakInteraction1985note\endcsname

\bibitem[{Goldreich \& Schubert(1967)}]{goldreichDifferentialRotationStars1967}
Goldreich, P. \& Schubert, G. 1967, The Astrophysical Journal, 150, 571 \csname
  goldreichDifferentialRotationStars1967link\endcsname~\csname
  goldreichDifferentialRotationStars1967note\endcsname

\bibitem[{Graczyk {et~al.}(2018)Graczyk, Pietrzy{\'n}ski, Thompson, Gieren,
  Pilecki, Konorski, Villanova, G{\'o}rski, Suchomska, Karczmarek, Stepie{\'n},
  Storm, Taormina, Ko{\l}aczkowski, Wielg{\'o}rski, Narloch, Zgirski, Gallenne,
  Ostrowski, Smolec, Udalski, Soszy{\'n}ski, Kervella, Nardetto, Szyma{\'n}ski,
  Wyrzykowski, Ulaczyk, Poleski, Pietrukowicz, Koz{\l}owski, Skowron, \&
  Mr{\'o}z}]{graczykLatetypeEclipsingBinaries2018}
Graczyk, D., Pietrzy{\'n}ski, G., Thompson, I.~B., {et~al.} 2018, The
  Astrophysical Journal, 860, 1 \csname
  graczykLatetypeEclipsingBinaries2018link\endcsname~\csname
  graczykLatetypeEclipsingBinaries2018note\endcsname

\bibitem[{Graczyk {et~al.}(2011)Graczyk, Soszy{\'n}ski, Poleski,
  Pietrzy{\'n}ski, Udalski, Szyma{\'n}ski, Kubiak, Wyrzykowski, \&
  Ulaczyk}]{graczykOpticalGravitationalLensing2011}
Graczyk, D., Soszy{\'n}ski, I., Poleski, R., {et~al.} 2011, Acta Astronomica,
  61, 103 \csname graczykOpticalGravitationalLensing2011link\endcsname~\csname
  graczykOpticalGravitationalLensing2011note\endcsname

\bibitem[{Guo {et~al.}(2022)Guo, Li, Xiong, Li, Wang, Xiong, Luo, Hou, Liu,
  Han, \& Chen}]{guoBinarityEarlytypeStars2022}
Guo, Y., Li, J., Xiong, J., {et~al.} 2022, Research in Astronomy and
  Astrophysics, 22, 025009 \csname
  guoBinarityEarlytypeStars2022link\endcsname~\csname
  guoBinarityEarlytypeStars2022note\endcsname

\bibitem[{Hainich {et~al.}(2014)Hainich, R{\"u}hling, Todt, Oskinova, Liermann,
  Gr{\"a}fener, Foellmi, Schnurr, \& Hamann}]{hainichWolfRayetStarsLarge2014}
Hainich, R., R{\"u}hling, U., Todt, H., {et~al.} 2014, Astronomy and
  Astrophysics, 565, A27 \csname
  hainichWolfRayetStarsLarge2014link\endcsname~\csname
  hainichWolfRayetStarsLarge2014note\endcsname

\bibitem[{Hamann {et~al.}(1995)Hamann, Koesterke, \&
  Wessolowski}]{hamannSpectralAnalysesGalactic1995}
Hamann, W.-R., Koesterke, L., \& Wessolowski, U. 1995, Astronomy \&
  Astrophysics, 299, 151 \csname
  hamannSpectralAnalysesGalactic1995link\endcsname~\csname
  hamannSpectralAnalysesGalactic1995note\endcsname

\bibitem[{Harris {et~al.}(2020)Harris, Millman, {van der Walt}, Gommers,
  Virtanen, Cournapeau, Wieser, Taylor, Berg, Smith, Kern, Picus, Hoyer, {van
  Kerkwijk}, Brett, Haldane, {del R{\'i}o}, Wiebe, Peterson,
  {G{\'e}rard-Marchant}, Sheppard, Reddy, Weckesser, Abbasi, Gohlke, \&
  Oliphant}]{harrisArrayProgrammingNumPy2020}
Harris, C.~R., Millman, K.~J., {van der Walt}, S.~J., {et~al.} 2020, Nature,
  585, 357 \csname harrisArrayProgrammingNumPy2020link\endcsname~\csname
  harrisArrayProgrammingNumPy2020note\endcsname

\bibitem[{Hazlehurst(1993)}]{hazlehurstEquilibriumContactBinary1993}
Hazlehurst, J. 1993, Astronomy \& Astrophysics, 271, 209 \csname
  hazlehurstEquilibriumContactBinary1993link\endcsname~\csname
  hazlehurstEquilibriumContactBinary1993note\endcsname

\bibitem[{Hazlehurst \& Refsdal(1980)}]{hazlehurstStabilityAgezeroContact1980}
Hazlehurst, J. \& Refsdal, S. 1980, Astronomy \& Astrophysics, 84, 200 \csname
  hazlehurstStabilityAgezeroContact1980link\endcsname~\csname
  hazlehurstStabilityAgezeroContact1980note\endcsname

\bibitem[{Heger {et~al.}(2000)Heger, Langer, \&
  Woosley}]{hegerPresupernovaEvolutionRotating2000}
Heger, A., Langer, N., \& Woosley, S.~E. 2000, The Astrophysical Journal, 528,
  368 \csname hegerPresupernovaEvolutionRotating2000link\endcsname~\csname
  hegerPresupernovaEvolutionRotating2000note\endcsname

\bibitem[{Henneco {et~al.}(2024{\natexlab{a}})Henneco, Schneider, Hekker, \&
  Aerts}]{hennecoMergerSeismologyDistinguishing2024}
Henneco, J., Schneider, F. R.~N., Hekker, S., \& Aerts, C. 2024{\natexlab{a}},
  Merger Seismology: Distinguishing Massive Merger Products from Genuine Single
  Stars Using Asteroseismology \csname
  hennecoMergerSeismologyDistinguishing2024link\endcsname~\csname
  hennecoMergerSeismologyDistinguishing2024note\endcsname

\bibitem[{Henneco {et~al.}(2024{\natexlab{b}})Henneco, Schneider, \&
  Laplace}]{hennecoContactTracingBinary2024}
Henneco, J., Schneider, F. R.~N., \& Laplace, E. 2024{\natexlab{b}}, Astronomy
  \& Astrophysics, 682, A169 \csname
  hennecoContactTracingBinary2024link\endcsname~\csname
  hennecoContactTracingBinary2024note\endcsname

\bibitem[{Hilditch \& Evans(1985)}]{hilditchMassiveNearcontactBinary1985}
Hilditch, R.~W. \& Evans, T.~L. 1985, Monthly Notices of the Royal Astronomical
  Society, 213, 75 \csname
  hilditchMassiveNearcontactBinary1985link\endcsname~\csname
  hilditchMassiveNearcontactBinary1985note\endcsname

\bibitem[{Hilditch {et~al.}(2005)Hilditch, Howarth, \&
  Harries}]{hilditchFortyEclipsingBinaries2005}
Hilditch, R.~W., Howarth, I.~D., \& Harries, T.~J. 2005, Monthly Notices of the
  Royal Astronomical Society, 357, 304 \csname
  hilditchFortyEclipsingBinaries2005link\endcsname~\csname
  hilditchFortyEclipsingBinaries2005note\endcsname

\bibitem[{Hunter(2007)}]{hunterMatplotlib2DGraphics2007}
Hunter, J.~D. 2007, Computing in Science and Engineering, 9, 90 \csname
  hunterMatplotlib2DGraphics2007link\endcsname~\csname
  hunterMatplotlib2DGraphics2007note\endcsname

\bibitem[{Hurley {et~al.}(2002)Hurley, Tout, \&
  Pols}]{hurleyEvolutionBinaryStars2002}
Hurley, J.~R., Tout, C.~A., \& Pols, O.~R. 2002, Monthly Notices of the Royal
  Astronomical Society, 329, 897 \csname
  hurleyEvolutionBinaryStars2002link\endcsname~\csname
  hurleyEvolutionBinaryStars2002note\endcsname

\bibitem[{Iglesias \& Rogers(1993)}]{iglesiasRadiativeOpacitiesCarbon1993}
Iglesias, C.~A. \& Rogers, F.~J. 1993, The Astrophysical Journal, 412, 752
  \csname iglesiasRadiativeOpacitiesCarbon1993link\endcsname~\csname
  iglesiasRadiativeOpacitiesCarbon1993note\endcsname

\bibitem[{Iglesias \& Rogers(1996)}]{iglesiasUpdatedOpalOpacities1996}
Iglesias, C.~A. \& Rogers, F.~J. 1996, The Astrophysical Journal, 464, 943
  \csname iglesiasUpdatedOpalOpacities1996link\endcsname~\csname
  iglesiasUpdatedOpalOpacities1996note\endcsname

\bibitem[{Itoh {et~al.}(1996)Itoh, Hayashi, Nishikawa, \&
  Kohyama}]{itohNeutrinoEnergyLoss1996}
Itoh, N., Hayashi, H., Nishikawa, A., \& Kohyama, Y. 1996, The Astrophysical
  Journal Supplement Series, 102, 411 \csname
  itohNeutrinoEnergyLoss1996link\endcsname~\csname
  itohNeutrinoEnergyLoss1996note\endcsname

\bibitem[{Jermyn {et~al.}(2023)Jermyn, Bauer, Schwab, Farmer, Ball, Bellinger,
  Dotter, Joyce, Marchant, Mombarg, Wolf, Sunny~Wong, Cinquegrana, Farrell,
  Smolec, Thoul, Cantiello, Herwig, Toloza, Bildsten, Townsend, \&
  Timmes}]{jermynModulesExperimentsStellar2023}
Jermyn, A.~S., Bauer, E.~B., Schwab, J., {et~al.} 2023, The Astrophysical
  Journal Supplement Series, 265, 15 \csname
  jermynModulesExperimentsStellar2023link\endcsname~\csname
  jermynModulesExperimentsStellar2023note\endcsname

\bibitem[{Jermyn {et~al.}(2021)Jermyn, Schwab, Bauer, Timmes, \&
  Potekhin}]{jermynSkyeDifferentiableEquation2021}
Jermyn, A.~S., Schwab, J., Bauer, E., Timmes, F.~X., \& Potekhin, A.~Y. 2021,
  The Astrophysical Journal, 913, 72 \csname
  jermynSkyeDifferentiableEquation2021link\endcsname~\csname
  jermynSkyeDifferentiableEquation2021note\endcsname

\bibitem[{Johnston {et~al.}(2024)Johnston, Michielsen, Anders, Renzo,
  Cantiello, Marchant, Goldberg, Townsend, Sabhahit, \&
  Jermyn}]{johnstonModellingTimedependentConvective2024}
Johnston, C., Michielsen, M., Anders, E.~H., {et~al.} 2024, The Astrophysical
  Journal, 964, 170 \csname
  johnstonModellingTimedependentConvective2024link\endcsname~\csname
  johnstonModellingTimedependentConvective2024note\endcsname

\bibitem[{K{\"a}hler(1989)}]{kahlerStructureEquationsContact1989}
K{\"a}hler, H. 1989, Astronomy \& Astrophysics, 209, 67 \csname
  kahlerStructureEquationsContact1989link\endcsname~\csname
  kahlerStructureEquationsContact1989note\endcsname

\bibitem[{Kippenhahn {et~al.}(1980)Kippenhahn, Ruschenplatt, \&
  Thomas}]{kippenhahnTimeScaleThermohaline1980}
Kippenhahn, R., Ruschenplatt, G., \& Thomas, H.~C. 1980, Astronomy \&
  Astrophysics, 91, 175 \csname
  kippenhahnTimeScaleThermohaline1980link\endcsname~\csname
  kippenhahnTimeScaleThermohaline1980note\endcsname

\bibitem[{Kobulnicky {et~al.}(2014)Kobulnicky, Kiminki, Lundquist, Burke,
  Chapman, Keller, Lester, Rolen, Topel, Bhattacharjee, Smullen, {\'A}lvarez,
  Runnoe, Dale, \& Brotherton}]{kobulnickyCOMPLETESTATISTICSMASSIVE2014}
Kobulnicky, H.~A., Kiminki, D.~C., Lundquist, M.~J., {et~al.} 2014, The
  Astrophysical Journal Supplement Series, 213, 34 \csname
  kobulnickyCOMPLETESTATISTICSMASSIVE2014link\endcsname~\csname
  kobulnickyCOMPLETESTATISTICSMASSIVE2014note\endcsname

\bibitem[{Kroupa(2001)}]{kroupaVariationInitialMass2001}
Kroupa, P. 2001, Monthly Notices of the Royal Astronomical Society, 322, 231
  \csname kroupaVariationInitialMass2001link\endcsname~\csname
  kroupaVariationInitialMass2001note\endcsname

\bibitem[{Langanke \&
  {Mart{\'i}nez-Pinedo}(2000)}]{langankeShellmodelCalculationsStellar2000}
Langanke, K. \& {Mart{\'i}nez-Pinedo}, G. 2000, Nuclear Physics A, 673, 481
  \csname langankeShellmodelCalculationsStellar2000link\endcsname~\csname
  langankeShellmodelCalculationsStellar2000note\endcsname

\bibitem[{Langer {et~al.}(1983)Langer, Fricke, \&
  Sugimoto}]{langerSemiconvectiveDiffusionEnergy1983}
Langer, N., Fricke, K.~J., \& Sugimoto, D. 1983, Astronomy \& Astrophysics,
  126, 207 \csname
  langerSemiconvectiveDiffusionEnergy1983link\endcsname~\csname
  langerSemiconvectiveDiffusionEnergy1983note\endcsname

\bibitem[{Lanthermann {et~al.}(2023)Lanthermann, Le~Bouquin, Sana, M{\'e}rand,
  Monnier, Perraut, Frost, Mahy, Gosset, De~Becker, Kraus, Anugu, Davies,
  Ennis, Gardner, Labdon, Setterholm, {ten Brummelaar}, \&
  Schaefer}]{lanthermannMultiplicityNorthernBright2023}
Lanthermann, C., Le~Bouquin, J.~B., Sana, H., {et~al.} 2023, Astronomy \&
  Astrophysics, 672, A6 \csname
  lanthermannMultiplicityNorthernBright2023link\endcsname~\csname
  lanthermannMultiplicityNorthernBright2023note\endcsname

\bibitem[{Ledoux(1947)}]{ledouxStellarModelsConvection1947}
Ledoux, P. 1947, The Astrophysical Journal, 105, 305 \csname
  ledouxStellarModelsConvection1947link\endcsname~\csname
  ledouxStellarModelsConvection1947note\endcsname

\bibitem[{Leung {et~al.}(1984)Leung, Sistero, Zhai, Grieco, \&
  Candellero}]{leungRevisedUBVPhotometric1984}
Leung, K.~C., Sistero, R.~F., Zhai, D.~S., Grieco, A., \& Candellero, B. 1984,
  The Astronomical Journal, 89, 872 \csname
  leungRevisedUBVPhotometric1984link\endcsname~\csname
  leungRevisedUBVPhotometric1984note\endcsname

\bibitem[{Li {et~al.}(2022{\natexlab{a}})Li, Liao, Qian,
  Fern{\'a}ndez~Laj{\'u}s, Zhang, \& Zhao}]{liV606CenNewly2022}
Li, F.~X., Liao, W.~P., Qian, S.~B., {et~al.} 2022{\natexlab{a}}, The
  Astrophysical Journal, 924, 30 \csname
  liV606CenNewly2022link\endcsname~\csname liV606CenNewly2022note\endcsname

\bibitem[{Li {et~al.}(2022{\natexlab{b}})Li, Qian, Jiao, \&
  Ma}]{liTwoMassiveClose2022}
Li, F.~X., Qian, S.~B., Jiao, C.~L., \& Ma, W.~W. 2022{\natexlab{b}}, The
  Astrophysical Journal, 932, 14 \csname
  liTwoMassiveClose2022link\endcsname~\csname
  liTwoMassiveClose2022note\endcsname

\bibitem[{Linnell \& Scheick(1991)}]{linnellDoesSVCentauri1991}
Linnell, A.~P. \& Scheick, X. 1991, The Astrophysical Journal, 379, 721 \csname
  linnellDoesSVCentauri1991link\endcsname~\csname
  linnellDoesSVCentauri1991note\endcsname

\bibitem[{Lorenz {et~al.}(1999)Lorenz, Mayer, \&
  Drechsel}]{lorenzV606CentauriEarlytype1999}
Lorenz, R., Mayer, P., \& Drechsel, H. 1999, Astronomy \& Astrophysics, 345,
  531 \csname lorenzV606CentauriEarlytype1999link\endcsname~\csname
  lorenzV606CentauriEarlytype1999note\endcsname

\bibitem[{Lorenzo {et~al.}(2014)Lorenzo, Negueruela, Baker, Garc{\'i}a,
  {Sim{\'o}n-D{\'i}az}, Pastor, \&
  M{\'e}ndez~Majuelos}]{lorenzoMYCamelopardalisVery2014}
Lorenzo, J., Negueruela, I., Baker, A. K. F.~V., {et~al.} 2014, Astronomy \&
  Astrophysics, 572, A110 \csname
  lorenzoMYCamelopardalisVery2014link\endcsname~\csname
  lorenzoMYCamelopardalisVery2014note\endcsname

\bibitem[{Lorenzo {et~al.}(2017)Lorenzo, {Sim{\'o}n-D{\'i}az}, Negueruela,
  Vilardell, Garcia, Evans, \& Montes}]{lorenzoMassiveMultipleSystem2017}
Lorenzo, J., {Sim{\'o}n-D{\'i}az}, S., Negueruela, I., {et~al.} 2017, Astronomy
  \& Astrophysics, 606, A54 \csname
  lorenzoMassiveMultipleSystem2017link\endcsname~\csname
  lorenzoMassiveMultipleSystem2017note\endcsname

\bibitem[{Lucy(1968)}]{lucyStructureContactBinaries1968}
Lucy, L.~B. 1968, The Astrophysical Journal, 151, 1123 \csname
  lucyStructureContactBinaries1968link\endcsname~\csname
  lucyStructureContactBinaries1968note\endcsname

\bibitem[{Mahy {et~al.}(2020{\natexlab{a}})Mahy, Almeida, Sana, Clark, {de
  Koter}, {de Mink}, Evans, Grin, Langer, Moffat, Schneider, Shenar, \&
  Tramper}]{mahyTarantulaMassiveBinary2020b}
Mahy, L., Almeida, L.~A., Sana, H., {et~al.} 2020{\natexlab{a}}, Astronomy \&
  Astrophysics, 634, A119 \csname
  mahyTarantulaMassiveBinary2020blink\endcsname~\csname
  mahyTarantulaMassiveBinary2020bnote\endcsname

\bibitem[{Mahy {et~al.}(2020{\natexlab{b}})Mahy, Sana, {Abdul-Masih}, Almeida,
  Langer, Shenar, {de Koter}, {de Mink}, {de Wit}, Grin, Evans, Moffat,
  Schneider, Barb{\'a}, Clark, Crowther, Gr{\"a}fener, Lennon, Tramper, \&
  Vink}]{mahyTarantulaMassiveBinary2020a}
Mahy, L., Sana, H., {Abdul-Masih}, M., {et~al.} 2020{\natexlab{b}}, Astronomy
  \& Astrophysics, 634, A118 \csname
  mahyTarantulaMassiveBinary2020alink\endcsname~\csname
  mahyTarantulaMassiveBinary2020anote\endcsname

\bibitem[{Mandel \& {de Mink}(2016)}]{mandelMergingBinaryBlack2016}
Mandel, I. \& {de Mink}, S.~E. 2016, Monthly Notices of the Royal Astronomical
  Society, 458, 2634 \csname mandelMergingBinaryBlack2016link\endcsname~\csname
  mandelMergingBinaryBlack2016note\endcsname

\bibitem[{Marchant \& Bodensteiner(2024)}]{marchantEvolutionMassiveBinary2024}
Marchant, P. \& Bodensteiner, J. 2024, Annual Review of Astronomy and
  Astrophysics, 62, 21 \csname
  marchantEvolutionMassiveBinary2024link\endcsname~\csname
  marchantEvolutionMassiveBinary2024note\endcsname

\bibitem[{Marchant {et~al.}(2016)Marchant, Langer, Podsiadlowski, Tauris, \&
  Moriya}]{marchantNewRouteMerging2016}
Marchant, P., Langer, N., Podsiadlowski, P., Tauris, T.~M., \& Moriya, T.~J.
  2016, Astronomy \& Astrophysics, 588, A50 \csname
  marchantNewRouteMerging2016link\endcsname~\csname
  marchantNewRouteMerging2016note\endcsname

\bibitem[{Marchant {et~al.}(2021)Marchant, Pappas, {Gallegos-Garcia}, Berry,
  Taam, Kalogera, \& Podsiadlowski}]{marchantRoleMassTransfer2021}
Marchant, P., Pappas, K. M.~W., {Gallegos-Garcia}, M., {et~al.} 2021, Astronomy
  \& Astrophysics, 650, A107 \csname
  marchantRoleMassTransfer2021link\endcsname~\csname
  marchantRoleMassTransfer2021note\endcsname

\bibitem[{Martins {et~al.}(2017)Martins, Mahy, \&
  Herv{\'e}}]{martinsPropertiesSixShortperiod2017}
Martins, F., Mahy, L., \& Herv{\'e}, A. 2017, Astronomy \& Astrophysics, 607,
  A82 \csname martinsPropertiesSixShortperiod2017link\endcsname~\csname
  martinsPropertiesSixShortperiod2017note\endcsname

\bibitem[{Mason {et~al.}(2009)Mason, Hartkopf, Gies, Henry, \&
  Helsel}]{masonHighAngularResolution2009}
Mason, B.~D., Hartkopf, W.~I., Gies, D.~R., Henry, T.~J., \& Helsel, J.~W.
  2009, The Astronomical Journal, 137, 3358 \csname
  masonHighAngularResolution2009link\endcsname~\csname
  masonHighAngularResolution2009note\endcsname

\bibitem[{Mayer {et~al.}(2013)Mayer, Drechsel, Harmanec, Yang, \& {\v
  S}lechta}]{mayerOtypeEclipsingContact2013}
Mayer, P., Drechsel, H., Harmanec, P., Yang, S., \& {\v S}lechta, M. 2013,
  Astronomy \& Astrophysics, 559, A22 \csname
  mayerOtypeEclipsingContact2013link\endcsname~\csname
  mayerOtypeEclipsingContact2013note\endcsname

\bibitem[{Menon {et~al.}(2021)Menon, Langer, {de Mink}, Justham, Sen,
  Sz{\'e}csi, {de Koter}, {Abdul-Masih}, Sana, Mahy, \&
  Marchant}]{menonDetailedEvolutionaryModels2021}
Menon, A., Langer, N., {de Mink}, S.~E., {et~al.} 2021, Monthly Notices of the
  Royal Astronomical Society, 507, 5013 \csname
  menonDetailedEvolutionaryModels2021link\endcsname~\csname
  menonDetailedEvolutionaryModels2021note\endcsname

\bibitem[{Menon {et~al.}(2024)Menon, Pawlak, Lennon, Sen, \&
  Langer}]{menonStudySixtyoneMassive2024}
Menon, A., Pawlak, M., Lennon, D.~J., Sen, K., \& Langer, N. 2024, A Study of
  Sixty-One Massive over-Contact Binary Candidates from the {{OGLE}} Survey of
  the {{Magellanic Clouds}} \csname
  menonStudySixtyoneMassive2024link\endcsname~\csname
  menonStudySixtyoneMassive2024note\endcsname

\bibitem[{Michielsen {et~al.}(2021)Michielsen, Aerts, \&
  Bowman}]{michielsenProbingTemperatureGradient2021}
Michielsen, M., Aerts, C., \& Bowman, D.~M. 2021, Astronomy and Astrophysics,
  650, A175 \csname
  michielsenProbingTemperatureGradient2021link\endcsname~\csname
  michielsenProbingTemperatureGradient2021note\endcsname

\bibitem[{Moe \& Di~Stefano(2017)}]{moeMindYourPs2017}
Moe, M. \& Di~Stefano, R. 2017, The Astrophysical Journal Supplement Series,
  230, 15 \csname moeMindYourPs2017link\endcsname~\csname
  moeMindYourPs2017note\endcsname

\bibitem[{Nelson \& Eggleton(2001)}]{nelsonCompleteSurveyCase2001}
Nelson, C.~A. \& Eggleton, P.~P. 2001, The Astrophysical Journal, 552, 664
  \csname nelsonCompleteSurveyCase2001link\endcsname~\csname
  nelsonCompleteSurveyCase2001note\endcsname

\bibitem[{Nieuwenhuijzen \& {de
  Jager}(1995)}]{nieuwenhuijzenAtmosphericAccelerationsStability1995}
Nieuwenhuijzen, H. \& {de Jager}, C. 1995, Astronomy \& Astrophysics, 302, 811
  \csname
  nieuwenhuijzenAtmosphericAccelerationsStability1995link\endcsname~\csname
  nieuwenhuijzenAtmosphericAccelerationsStability1995note\endcsname

\bibitem[{Oda {et~al.}(1994)Oda, Hino, Muto, Takahara, \&
  Sato}]{odaRateTablesWeak1994}
Oda, T., Hino, M., Muto, K., Takahara, M., \& Sato, K. 1994, Atomic Data and
  Nuclear Data Tables, 56, 231 \csname
  odaRateTablesWeak1994link\endcsname~\csname
  odaRateTablesWeak1994note\endcsname

\bibitem[{{\"O}pik(1924)}]{opikStatisticalStudiesDouble1924}
{\"O}pik, E. 1924, Publications of the Tartu Astrofizica Observatory, 25, 1
  \csname opikStatisticalStudiesDouble1924link\endcsname~\csname
  opikStatisticalStudiesDouble1924note\endcsname

\bibitem[{Ostrov(2001)}]{ostrovOrbitalSolutionMACHO2001}
Ostrov, P.~G. 2001, Monthly Notices of the Royal Astronomical Society, 321, L25
  \csname ostrovOrbitalSolutionMACHO2001link\endcsname~\csname
  ostrovOrbitalSolutionMACHO2001note\endcsname

\bibitem[{Paczynski {et~al.}(2006)Paczynski, Szczygiel, Pilecki, \&
  Pojmanski}]{paczynskiEclipsingBinariesASAS2006}
Paczynski, B., Szczygiel, D., Pilecki, B., \& Pojmanski, G. 2006, Monthly
  Notices of the Royal Astronomical Society, 368, 1311 \csname
  paczynskiEclipsingBinariesASAS2006link\endcsname~\csname
  paczynskiEclipsingBinariesASAS2006note\endcsname

\bibitem[{Papaloizou \&
  Pringle(1979)}]{papaloizouMaintenanceTemperatureDiscontinuity1979}
Papaloizou, J. \& Pringle, J.~E. 1979, Monthly Notices of the Royal
  Astronomical Society, 189, 5P \csname
  papaloizouMaintenanceTemperatureDiscontinuity1979link\endcsname~\csname
  papaloizouMaintenanceTemperatureDiscontinuity1979note\endcsname

\bibitem[{Pauli {et~al.}(2022)Pauli, Langer, {Aguilera-Dena}, Wang, \&
  Marchant}]{pauliSyntheticPopulationWolfRayet2022}
Pauli, D., Langer, N., {Aguilera-Dena}, D.~R., Wang, C., \& Marchant, P. 2022,
  Astronomy \& Astrophysics, 667, A58 \csname
  pauliSyntheticPopulationWolfRayet2022link\endcsname~\csname
  pauliSyntheticPopulationWolfRayet2022note\endcsname

\bibitem[{Paxton {et~al.}(2011)Paxton, Bildsten, Dotter, Herwig, Lesaffre, \&
  Timmes}]{paxtonModulesExperimentsStellar2011}
Paxton, B., Bildsten, L., Dotter, A., {et~al.} 2011, The Astrophysical Journal
  Supplement Series, 192, 3 \csname
  paxtonModulesExperimentsStellar2011link\endcsname~\csname
  paxtonModulesExperimentsStellar2011note\endcsname

\bibitem[{Paxton {et~al.}(2013)Paxton, Cantiello, Arras, Bildsten, Brown,
  Dotter, Mankovich, Montgomery, Stello, Timmes, \&
  Townsend}]{paxtonModulesExperimentsStellar2013}
Paxton, B., Cantiello, M., Arras, P., {et~al.} 2013, The Astrophysical Journal
  Supplement Series, 208, 4 \csname
  paxtonModulesExperimentsStellar2013link\endcsname~\csname
  paxtonModulesExperimentsStellar2013note\endcsname

\bibitem[{Paxton {et~al.}(2015)Paxton, Marchant, Schwab, Bauer, Bildsten,
  Cantiello, Dessart, Farmer, Hu, Langer, Townsend, Townsley, \&
  Timmes}]{paxtonModulesExperimentsStellar2015}
Paxton, B., Marchant, P., Schwab, J., {et~al.} 2015, The Astrophysical Journal
  Supplement Series, 220, 15 \csname
  paxtonModulesExperimentsStellar2015link\endcsname~\csname
  paxtonModulesExperimentsStellar2015note\endcsname

\bibitem[{Paxton {et~al.}(2018)Paxton, Schwab, Bauer, Bildsten, Blinnikov,
  Duffell, Farmer, Goldberg, Marchant, Sorokina, Thoul, Townsend, \&
  Timmes}]{paxtonModulesExperimentsStellar2018}
Paxton, B., Schwab, J., Bauer, E.~B., {et~al.} 2018, The Astrophysical Journal
  Supplement Series, 234, 34 \csname
  paxtonModulesExperimentsStellar2018link\endcsname~\csname
  paxtonModulesExperimentsStellar2018note\endcsname

\bibitem[{Paxton {et~al.}(2019)Paxton, Smolec, Schwab, Gautschy, Bildsten,
  Cantiello, Dotter, Farmer, Goldberg, Jermyn, Kanbur, Marchant, Thoul,
  Townsend, Wolf, Zhang, \& Timmes}]{paxtonModulesExperimentsStellar2019}
Paxton, B., Smolec, R., Schwab, J., {et~al.} 2019, The Astrophysical Journal
  Supplement Series, 243, 10 \csname
  paxtonModulesExperimentsStellar2019link\endcsname~\csname
  paxtonModulesExperimentsStellar2019note\endcsname

\bibitem[{Pedersen {et~al.}(2021)Pedersen, Aerts, P{\'a}pics, Michielsen,
  Gebruers, Rogers, Molenberghs, Burssens, Garcia, \&
  Bowman}]{pedersenInternalMixingRotating2021}
Pedersen, M.~G., Aerts, C., P{\'a}pics, P.~I., {et~al.} 2021, Nature Astronomy,
  5, 715 \csname pedersenInternalMixingRotating2021link\endcsname~\csname
  pedersenInternalMixingRotating2021note\endcsname

\bibitem[{Penny {et~al.}(2008)Penny, Ouzts, \&
  Gies}]{pennyTomographicSeparationComposite2008}
Penny, L.~R., Ouzts, C., \& Gies, D.~R. 2008, The Astrophysical Journal, 681,
  554 \csname pennyTomographicSeparationComposite2008link\endcsname~\csname
  pennyTomographicSeparationComposite2008note\endcsname

\bibitem[{Pols(1994)}]{polsCaseEvolutionMassive1994}
Pols, O.~R. 1994, Astronomy \& Astrophysics, 290, 119 \csname
  polsCaseEvolutionMassive1994link\endcsname~\csname
  polsCaseEvolutionMassive1994note\endcsname

\bibitem[{Polushina(2004)}]{polushinaCatalogueMassiveClose2004}
Polushina, T.~S. 2004, Astronomical and Astrophysical Transactions, 23, 213
  \csname polushinaCatalogueMassiveClose2004link\endcsname~\csname
  polushinaCatalogueMassiveClose2004note\endcsname

\bibitem[{Potekhin \& Chabrier(2010)}]{potekhinThermodynamicFunctionsDense2010}
Potekhin, A.~Y. \& Chabrier, G. 2010, Contributions to Plasma Physics, 50, 82
  \csname potekhinThermodynamicFunctionsDense2010link\endcsname~\csname
  potekhinThermodynamicFunctionsDense2010note\endcsname

\bibitem[{Poutanen(2017)}]{poutanenRosselandFluxMean2017}
Poutanen, J. 2017, The Astrophysical Journal, 835, 119 \csname
  poutanenRosselandFluxMean2017link\endcsname~\csname
  poutanenRosselandFluxMean2017note\endcsname

\bibitem[{{Ram{\'i}rez-Tannus} {et~al.}(2021){Ram{\'i}rez-Tannus}, Backs, {de
  Koter}, Sana, Beuther, Bik, Brandner, Kaper, Linz, Henning, \&
  Poorta}]{ramirez-tannusRelationRadialVelocity2021}
{Ram{\'i}rez-Tannus}, M.~C., Backs, F., {de Koter}, A., {et~al.} 2021,
  Astronomy \& Astrophysics, 645, L10 \csname
  ramirez-tannusRelationRadialVelocity2021link\endcsname~\csname
  ramirez-tannusRelationRadialVelocity2021note\endcsname

\bibitem[{{Ram{\'i}rez-Tannus} {et~al.}(2024){Ram{\'i}rez-Tannus}, Derkink,
  Backs, {de Koter}, Sana, Poorta, Kaper, \&
  Stoop}]{ramirez-tannusSpectroscopicBinaryFraction2024}
{Ram{\'i}rez-Tannus}, M.~C., Derkink, A.~R., Backs, F., {et~al.} 2024,
  Astronomy and Astrophysics, 690, A178 \csname
  ramirez-tannusSpectroscopicBinaryFraction2024link\endcsname~\csname
  ramirez-tannusSpectroscopicBinaryFraction2024note\endcsname

\bibitem[{Raucq {et~al.}(2017)Raucq, Gosset, Rauw, Manfroid, Mahy, Mennekens,
  \& Vanbeveren}]{raucqObservationalSignaturesMassexchange2017}
Raucq, F., Gosset, E., Rauw, G., {et~al.} 2017, Astronomy \& Astrophysics, 601,
  A133 \csname
  raucqObservationalSignaturesMassexchange2017link\endcsname~\csname
  raucqObservationalSignaturesMassexchange2017note\endcsname

\bibitem[{Rogers \& Nayfonov(2002)}]{rogersUpdatedExpandedOPAL2002}
Rogers, F.~J. \& Nayfonov, A. 2002, The Astrophysical Journal, 576, 1064
  \csname rogersUpdatedExpandedOPAL2002link\endcsname~\csname
  rogersUpdatedExpandedOPAL2002note\endcsname

\bibitem[{Salpeter(1955)}]{salpeterLuminosityFunctionStellar1955}
Salpeter, E.~E. 1955, The Astrophysical Journal, 121, 161 \csname
  salpeterLuminosityFunctionStellar1955link\endcsname~\csname
  salpeterLuminosityFunctionStellar1955note\endcsname

\bibitem[{Sana {et~al.}(2013)Sana, {de Koter}, {de Mink}, Dunstall, Evans,
  {H{\'e}nault-Brunet}, Ma{\'i}z~Apell{\'a}niz, {Ram{\'i}rez-Agudelo}, Taylor,
  Walborn, Clark, Crowther, Herrero, Gieles, Langer, Lennon, \&
  Vink}]{sanaVLTFLAMESTarantulaSurvey2013}
Sana, H., {de Koter}, A., {de Mink}, S.~E., {et~al.} 2013, Astronomy \&
  Astrophysics, 550, A107 \csname
  sanaVLTFLAMESTarantulaSurvey2013link\endcsname~\csname
  sanaVLTFLAMESTarantulaSurvey2013note\endcsname

\bibitem[{Sana {et~al.}(2012)Sana, de~Mink, de~Koter, Langer, Evans, Gieles,
  Gosset, Izzard, Bouquin, \& Schneider}]{sanaBinaryInteractionDominates2012}
Sana, H., de~Mink, S.~E., de~Koter, A., {et~al.} 2012, Science, 337, 444
  \csname sanaBinaryInteractionDominates2012link\endcsname~\csname
  sanaBinaryInteractionDominates2012note\endcsname

\bibitem[{Sana {et~al.}(2014)Sana, Le~Bouquin, Lacour, Berger, Duvert, Gauchet,
  Norris, Olofsson, Pickel, Zins, Absil, {de Koter}, Kratter, Schnurr, \&
  Zinnecker}]{sanaSouthernMassiveStars2014}
Sana, H., Le~Bouquin, J.~B., Lacour, S., {et~al.} 2014, The Astrophysical
  Journal Supplement Series, 215, 15 \csname
  sanaSouthernMassiveStars2014link\endcsname~\csname
  sanaSouthernMassiveStars2014note\endcsname

\bibitem[{Sana {et~al.}(2017)Sana, {Ram{\'i}rez-Tannus}, {de Koter}, Kaper,
  Tramper, \& Bik}]{sanaDearthShortperiodMassive2017}
Sana, H., {Ram{\'i}rez-Tannus}, M.~C., {de Koter}, A., {et~al.} 2017, Astronomy
  \& Astrophysics, 599, L9 \csname
  sanaDearthShortperiodMassive2017link\endcsname~\csname
  sanaDearthShortperiodMassive2017note\endcsname

\bibitem[{Saumon {et~al.}(1995)Saumon, Chabrier, \& {van
  Horn}}]{saumonEquationStateLowMass1995}
Saumon, D., Chabrier, G., \& {van Horn}, H.~M. 1995, The Astrophysical Journal
  Supplement Series, 99, 713 \csname
  saumonEquationStateLowMass1995link\endcsname~\csname
  saumonEquationStateLowMass1995note\endcsname

\bibitem[{Schneider {et~al.}(2019)Schneider, Ohlmann, Podsiadlowski, R{\"o}pke,
  Balbus, Pakmor, \& Springel}]{schneiderStellarMergersOrigin2019}
Schneider, F. R.~N., Ohlmann, S.~T., Podsiadlowski, P., {et~al.} 2019, Nature,
  574, 211 \csname schneiderStellarMergersOrigin2019link\endcsname~\csname
  schneiderStellarMergersOrigin2019note\endcsname

\bibitem[{Selam(2004)}]{selamKeyParametersUMatype2004}
Selam, S.~O. 2004, Astronomy \& Astrophysics, 416, 1097 \csname
  selamKeyParametersUMatype2004link\endcsname~\csname
  selamKeyParametersUMatype2004note\endcsname

\bibitem[{Sen {et~al.}(2022)Sen, Langer, Marchant, Menon, {de Mink},
  Schootemeijer, Sch{\"u}rmann, Mahy, Hastings, Nathaniel, Sana, Wang, \&
  Xu}]{senDetailedModelsInteracting2022}
Sen, K., Langer, N., Marchant, P., {et~al.} 2022, Astronomy \& Astrophysics,
  659, A98 \csname senDetailedModelsInteracting2022link\endcsname~\csname
  senDetailedModelsInteracting2022note\endcsname

\bibitem[{Shenar {et~al.}(2020)Shenar, Bodensteiner, {Abdul-Masih}, Fabry,
  Mahy, Marchant, Banyard, Bowman, Dsilva, Hawcroft, Reggiani, \&
  Sana}]{shenarHiddenCompanionLB12020}
Shenar, T., Bodensteiner, J., {Abdul-Masih}, M., {et~al.} 2020, Astronomy \&
  Astrophysics, 639, L6 \csname
  shenarHiddenCompanionLB12020link\endcsname~\csname
  shenarHiddenCompanionLB12020note\endcsname

\bibitem[{Shenar {et~al.}(2022)Shenar, Sana, Mahy, Ma{\'i}z~Apell{\'a}niz,
  Crowther, Gromadzki, Herrero, Langer, Marchant, Schneider, Sen,
  Soszy{\'n}ski, \& Toonen}]{shenarTarantulaMassiveBinary2022}
Shenar, T., Sana, H., Mahy, L., {et~al.} 2022, Astronomy and Astrophysics, 665,
  A148 \csname shenarTarantulaMassiveBinary2022link\endcsname~\csname
  shenarTarantulaMassiveBinary2022note\endcsname

\bibitem[{Shu {et~al.}(1976)Shu, Lubow, \&
  Anderson}]{shuStructureContactBinaries1976}
Shu, F.~H., Lubow, S.~H., \& Anderson, L. 1976, The Astrophysical Journal, 209,
  536 \csname shuStructureContactBinaries1976link\endcsname~\csname
  shuStructureContactBinaries1976note\endcsname

\bibitem[{Shu {et~al.}(1979)Shu, Lubow, \&
  Anderson}]{shuStructureContactBinaries1979}
Shu, F.~H., Lubow, S.~H., \& Anderson, L. 1979, The Astrophysical Journal, 229,
  223 \csname shuStructureContactBinaries1979link\endcsname~\csname
  shuStructureContactBinaries1979note\endcsname

\bibitem[{Sweet(1950)}]{sweetImportanceRotationStellar1950}
Sweet, P.~A. 1950, Monthly Notices of the Royal Astronomical Society, 110, 548
  \csname sweetImportanceRotationStellar1950link\endcsname~\csname
  sweetImportanceRotationStellar1950note\endcsname

\bibitem[{Szymanski {et~al.}(2001)Szymanski, Kubiak, \&
  Udalski}]{szymanskiContactBinariesOGLEI2001}
Szymanski, M., Kubiak, M., \& Udalski, A. 2001, Acta Astronomica, 51, 259
  \csname szymanskiContactBinariesOGLEI2001link\endcsname~\csname
  szymanskiContactBinariesOGLEI2001note\endcsname

\bibitem[{Timmes \& Swesty(2000)}]{timmesAccuracyConsistencySpeed2000}
Timmes, F.~X. \& Swesty, F.~D. 2000, The Astrophysical Journal Supplement
  Series, 126, 501 \csname
  timmesAccuracyConsistencySpeed2000link\endcsname~\csname
  timmesAccuracyConsistencySpeed2000note\endcsname

\bibitem[{Tokovinin \& Moe(2020)}]{tokovininFormationCloseBinaries2020}
Tokovinin, A. \& Moe, M. 2020, Monthly Notices of the Royal Astronomical
  Society, 491, 5158 \csname
  tokovininFormationCloseBinaries2020link\endcsname~\csname
  tokovininFormationCloseBinaries2020note\endcsname

\bibitem[{Vanbeveren {et~al.}(1998)Vanbeveren, De~Donder, Van~Bever,
  Van~Rensbergen, \& De~Loore}]{vanbeverenWROtypeStar1998}
Vanbeveren, D., De~Donder, E., Van~Bever, J., Van~Rensbergen, W., \& De~Loore,
  C. 1998, New Astronomy, 3, 443 \csname
  vanbeverenWROtypeStar1998link\endcsname~\csname
  vanbeverenWROtypeStar1998note\endcsname

\bibitem[{Villase{\~n}or {et~al.}(2021)Villase{\~n}or, Taylor, Evans,
  {Ram{\'i}rez-Agudelo}, Sana, Almeida, {de Mink}, Dufton, \&
  Langer}]{villasenorBtypeBinariesCharacterization2021}
Villase{\~n}or, J.~I., Taylor, W.~D., Evans, C.~J., {et~al.} 2021, Monthly
  Notices of the Royal Astronomical Society, 507, 5348 \csname
  villasenorBtypeBinariesCharacterization2021link\endcsname~\csname
  villasenorBtypeBinariesCharacterization2021note\endcsname

\bibitem[{Vink {et~al.}(2001)Vink, {de Koter}, \&
  Lamers}]{vinkMasslossPredictionsStars2001}
Vink, J.~S., {de Koter}, A., \& Lamers, H. J. G. L.~M. 2001, Astronomy \&
  Astrophysics, 369, 574 \csname
  vinkMasslossPredictionsStars2001link\endcsname~\csname
  vinkMasslossPredictionsStars2001note\endcsname

\bibitem[{Virtanen {et~al.}(2020)Virtanen, Gommers, Oliphant, Haberland, Reddy,
  Cournapeau, Burovski, Peterson, Weckesser, Bright, {van der Walt}, Brett,
  Wilson, Millman, Mayorov, Nelson, Jones, Kern, Larson, Carey, Polat, Feng,
  Moore, VanderPlas, Laxalde, Perktold, Cimrman, Henriksen, Quintero, Harris,
  Archibald, Ribeiro, Pedregosa, {van Mulbregt}, \& {SciPy 1. 0
  Contributors}}]{virtanenSciPyFundamentalAlgorithms2020}
Virtanen, P., Gommers, R., Oliphant, T.~E., {et~al.} 2020, Nature Methods, 17,
  261 \csname virtanenSciPyFundamentalAlgorithms2020link\endcsname~\csname
  virtanenSciPyFundamentalAlgorithms2020note\endcsname

\bibitem[{Vitense(1953)}]{vitenseWasserstoffkonvektionszoneSonneMit1953}
Vitense, E. 1953, Zeitschrift fur Astrophysik, 32, 135 \csname
  vitenseWasserstoffkonvektionszoneSonneMit1953link\endcsname~\csname
  vitenseWasserstoffkonvektionszoneSonneMit1953note\endcsname

\bibitem[{{von Zeipel}(1924)}]{vonzeipelRadiativeEquilibriumRotating1924}
{von Zeipel}, H. 1924, Monthly Notices of the Royal Astronomical Society, 84,
  665 \csname vonzeipelRadiativeEquilibriumRotating1924link\endcsname~\csname
  vonzeipelRadiativeEquilibriumRotating1924note\endcsname

\bibitem[{Vrancken {et~al.}(2024)Vrancken, {Abdul-Masih}, Escorza, Menon, Mahy,
  \& Marchant}]{vranckenConstrainingOvercontactPhase2024}
Vrancken, J., {Abdul-Masih}, M., Escorza, A., {et~al.} 2024, Constraining the
  Overcontact Phase in Massive Binary Evolution -- {{III}}. {{Period}}
  Stability of Known {{B}}+{{B}} and {{O}}+{{B}} Overcontact Systems \csname
  vranckenConstrainingOvercontactPhase2024link\endcsname~\csname
  vranckenConstrainingOvercontactPhase2024note\endcsname

\bibitem[{Wellstein {et~al.}(2001)Wellstein, Langer, \&
  Braun}]{wellsteinFormationContactMassive2001}
Wellstein, S., Langer, N., \& Braun, H. 2001, Astronomy and Astrophysics, 369,
  939 \csname wellsteinFormationContactMassive2001link\endcsname~\csname
  wellsteinFormationContactMassive2001note\endcsname

\bibitem[{Wilson \& Devinney(1971)}]{wilsonRealizationAccurateCloseBinary1971}
Wilson, R.~E. \& Devinney, E.~J. 1971, The Astrophysical Journal, 166, 605
  \csname wilsonRealizationAccurateCloseBinary1971link\endcsname~\csname
  wilsonRealizationAccurateCloseBinary1971note\endcsname

\bibitem[{Yang {et~al.}(2019)Yang, Yuan, \&
  Dai}]{yangComprehensiveStudyThree2019}
Yang, Y., Yuan, H., \& Dai, H. 2019, The Astronomical Journal, 157, 111 \csname
  yangComprehensiveStudyThree2019link\endcsname~\csname
  yangComprehensiveStudyThree2019note\endcsname

\bibitem[{Ya{\c s}arsoy \&
  Yakut(2014)}]{yasarsoyInteractingSupergiantClose2014}
Ya{\c s}arsoy, B. \& Yakut, K. 2014, New Astronomy, 31, 32 \csname
  yasarsoyInteractingSupergiantClose2014link\endcsname~\csname
  yasarsoyInteractingSupergiantClose2014note\endcsname

\bibitem[{Yoon \& Langer(2005)}]{yoonEvolutionRapidlyRotating2005}
Yoon, S.~C. \& Langer, N. 2005, Astronomy \& Astrophysics, 443, 643 \csname
  yoonEvolutionRapidlyRotating2005link\endcsname~\csname
  yoonEvolutionRapidlyRotating2005note\endcsname

\bibitem[{Yoon {et~al.}(2006)Yoon, Langer, \&
  Norman}]{yoonSingleStarProgenitors2006}
Yoon, S.-C., Langer, N., \& Norman, C. 2006, Astronomy \& Astrophysics, 460,
  199 \csname yoonSingleStarProgenitors2006link\endcsname~\csname
  yoonSingleStarProgenitors2006note\endcsname

\bibitem[{Zahn(1975)}]{zahnDynamicalTideClose1975}
Zahn, J.-P. 1975, Astronomy \& Astrophysics, 41, 329 \csname
  zahnDynamicalTideClose1975link\endcsname~\csname
  zahnDynamicalTideClose1975note\endcsname

\bibitem[{Zahn(1977)}]{zahnTidalFrictionClose1977}
Zahn, J.-P. 1977, Astronomy \& Astrophysics, 500, 121 \csname
  zahnTidalFrictionClose1977link\endcsname~\csname
  zahnTidalFrictionClose1977note\endcsname

\bibitem[{Zahn(1992)}]{zahnCirculationTurbulenceRotating1992}
Zahn, J.-P. 1992, Astronomy \& Astrophysics, 265, 115 \csname
  zahnCirculationTurbulenceRotating1992link\endcsname~\csname
  zahnCirculationTurbulenceRotating1992note\endcsname

\end{thebibliography}
\begin{appendix}
\section{Roche-lobe overflow at the zero-age main sequence} \label{app:pms_interaction}
In this appendix, we discuss the models that experience RLOF at the ZAMS.
These are marked with the backward hashing (\textbackslash\textbackslash) in Figs.~\ref{fig:et_completion}-\ref{fig:single_rot_completion}.
We expect these systems to strongly interact on the PMS, and to account for this possibility, we perform an artificial shrinking of the orbit toward their initial period.
We model these systems as follows.
\par

We start with the model at ZAMS as described in Sect~\ref{ssec:binary_init}.
In the case considered here, the more massive star will overflow the RL (potentially by a large amount since we ignore MT in the relaxation phase).
We then artificially modify the orbital period to put the components in a detached configuration, adding a certain amount of angular momentum to the orbit.
Finally, on a thermal timescale of the more massive star ($\tau = 0.75 GM_1^2/R_1L_1$), we drain the added angular momentum from the system to again reach the initial period, while allowing MT and ET.
This process allows the models to smoothly start RLOF, and is more of a technical modification to the models instead of being physically motivated.
A more accurate model would be to follow the evolution of forming massive stars throughout the PMS, and include some prescription of angular momentum loss through interaction with a disk or their natal cloud, but this is beyond the scope of this work.
\par

\subsection{Results}
\begin{figure}
	\centering
	\includegraphics{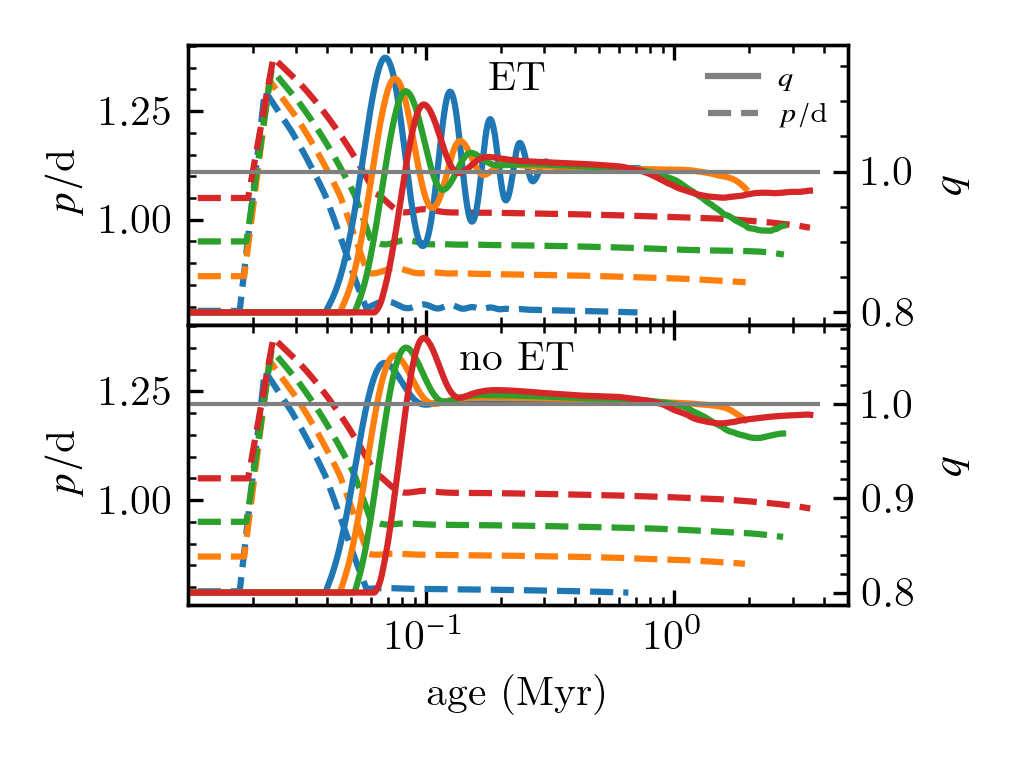}
	\caption{
		Period and mass-ratio evolution of the $M_{1, \rm init} = \qty{20}{\Msun}$ models with $q_{\rm init} = 0.8$ that experience RLOF at the ZAMS.
		The initial period of these models can be read off from the intersection with the left $y$-axis.
		All models interact before reaching their assumed initial period, meaning strong interaction on the PMS is expected.
	}
	\label{fig:zams_rlof}
\end{figure}

Figure \ref{fig:zams_rlof} shows the orbital period and mass-ratio evolution of the $\qty{20}{\Msun}$ model with initial mass ratio of $q=0.8$ and initial periods between $\qtyrange{0.8}{1.05}{\day}$.
Note the logarithmic scale on the $x$-axis.
The models reach our definition of the ZAMS at around $\qty{0.02}{\mega\year}$, after which we boost the period and initiate the shrinking on the thermal timescale of the primary.
\par

We emphasize that we do not treat the shrinking process fully.
\citet{sanaDearthShortperiodMassive2017} analyzed the RV-dispersion of young stars in M17, and inferred that the underlying period distribution fits best if there is a short-period cutoff at $p\approx \qty{50}{\day}$.
Presumably then, the stars start their migration inward toward a distribution that follows an \"Opik-like law, Eq.~\eqref{eq:p-dist}, down to periods of around $\qty{1}{\day}$.
Our toy migration model thus only covers the final 1\% of the migration into a RL filling configuration (from $\unsim \qty{1.2}{\day}$ to $\lesssim\!\!\qty{1}{\day}$).
It is however uncertain how, and on what timescale, migration happens, but the RV-dispersion analysis of \citet{ramirez-tannusRelationRadialVelocity2021} suggests it might be on the $\unit{Myr}$ nuclear timescale of massive stars.
Therefore, there is a possibility that short-period binaries evolve significantly before they are observed as $\unsim\qty{1}{\day}$ contact binaries.
In contrast, we operate the shrinking on a (short) thermal timescale, so that the two stellar models do not have time to build a significant composition difference before they enter contact.
\par

During the shrinking process, all models start MT before they reach their initial period again, signaling that significant interaction on the PMS could be important for very short-period binary systems.
We note that, with or without ET, all models equalize to mass ratios close to unity within the first $\qty{0.1}{\mega\year}$ or so.
The ET models show some oscillations because ET is actively regulating MT.
In cases where the energy donor is the mass accretor, MT is invigorated, and this is generally the case in thermal-timescale MT events as the accretor gains significant luminosity as the accreted material thermally relaxes.
Finally, it has not yet been unambiguously shown theoretically that homogeneous stellar models allow for a contact configuration with unequal masses.
While \citet{shuStructureContactBinaries1979} argue for a thermally stable structure with a temperature inversion layer, \citet{papaloizouMaintenanceTemperatureDiscontinuity1979} and \citet{hazlehurstStabilityAgezeroContact1980} show this violates the second law of thermodynamics, and the discontinuity disappears within a thermal timescale.
A model without such inconsistencies has not yet been presented.
Since we chose to perform the tidal shrinking on the thermal timescale, preventing composition gradients from building, our (numerical) evolution models, with or without ET, indicate that a composition difference is needed to create stable contact binaries with mass ratios away from unity.
\par

\begin{figure*}
	\centering
	\includegraphics[width=\textwidth]{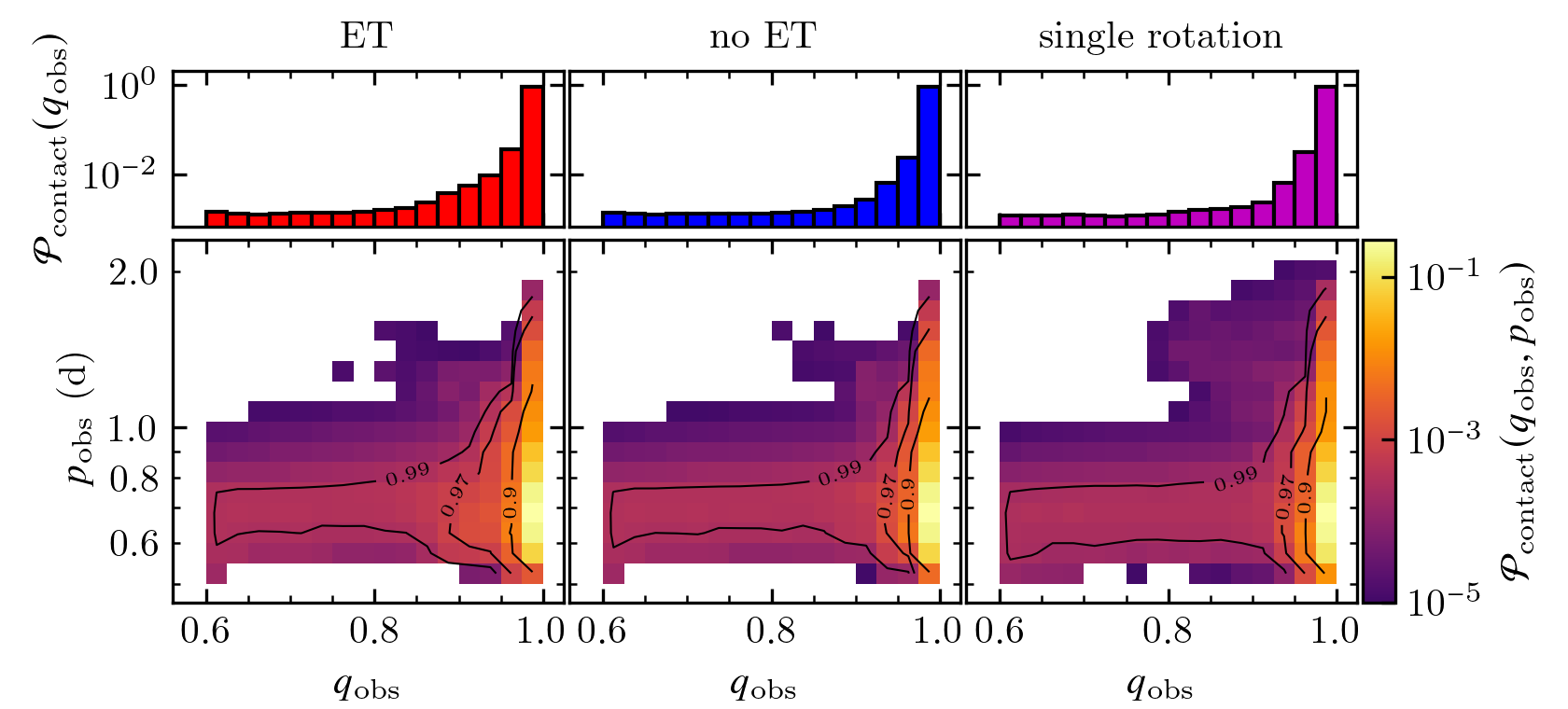}
	\caption{Probability density in $(q_{\rm obs}, p_{\rm obs})$ (lower panels) and $q_{\rm obs}$ space (upper panels) for the models that experience RLOF at the ZAMS.
		The thin black lines are highest density intervals of the indicated percentiles, i.e., we expect to observe at most 3\% of systems with a mass ratio significantly away from unity.}
	\label{fig:rlof_at_zams_dist}
\end{figure*}

\begin{figure}
	\centering
	\includegraphics{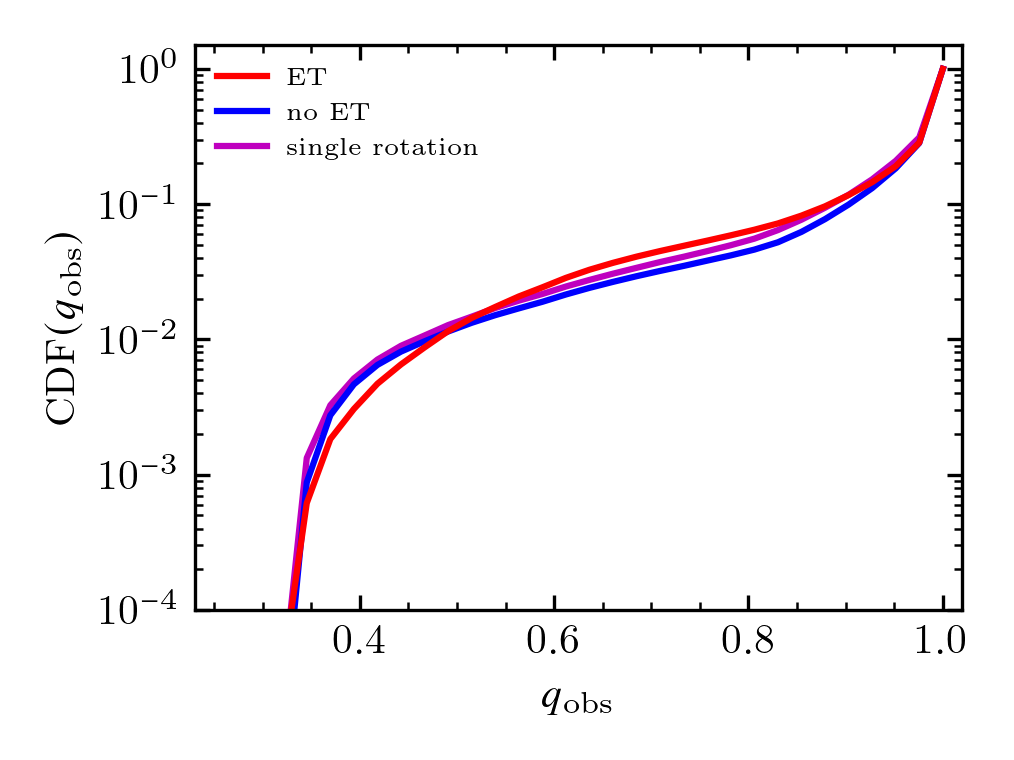}
	\caption{Cumulative probability distribution of the mass ratio of contact binaries that interact on the PMS.}
	\label{fig:zams_cumul}
\end{figure}

Despite some qualitative differences in their evolution, the ET and no-ET models show no significant difference in the mass-ratio distribution of contact binaries that experience RLOF at the ZAMS.
From Fig.~\ref{fig:rlof_at_zams_dist} and \ref{fig:zams_cumul}, we infer it is very unlikely to observe a contact binary that interacted on the PMS at mass ratios far from unity.
From our models, there is only around a 1\% chance a binary with a mass ratio more extreme than $q=0.8$ will be observed.
Conversely, over 95\% of all PMS-interacting binaries should be observed with $q>0.95$.
\par


Both observationally \citep[][]{sanaDearthShortperiodMassive2017, ramirez-tannusRelationRadialVelocity2021, ramirez-tannusSpectroscopicBinaryFraction2024} and theoretically \citep{tokovininFormationCloseBinaries2020}, it has been suggested that hardening of binaries through disk migration is an ingredient in the formation of close binaries, such as contact-binary progenitors.
Therefore, we think it fruitful to perform a population study of binaries with initial periods (on the PMS) of tens of days that harden over a $\unit{\mega\year}$ timescale.
\par
\FloatBarrier \section{Outcomes of the full grids}
In addition to the parameter space plot shown in Fig.~\ref{fig:completion}, Figs.~\ref{fig:et_completion}-\ref{fig:single_rot_completion} include the outcome of all simulations performed under the assumptions for energy transfer and tidal corrections.
\begin{sidewaysfigure*}
	\centering
	\includegraphics[width=\textwidth]{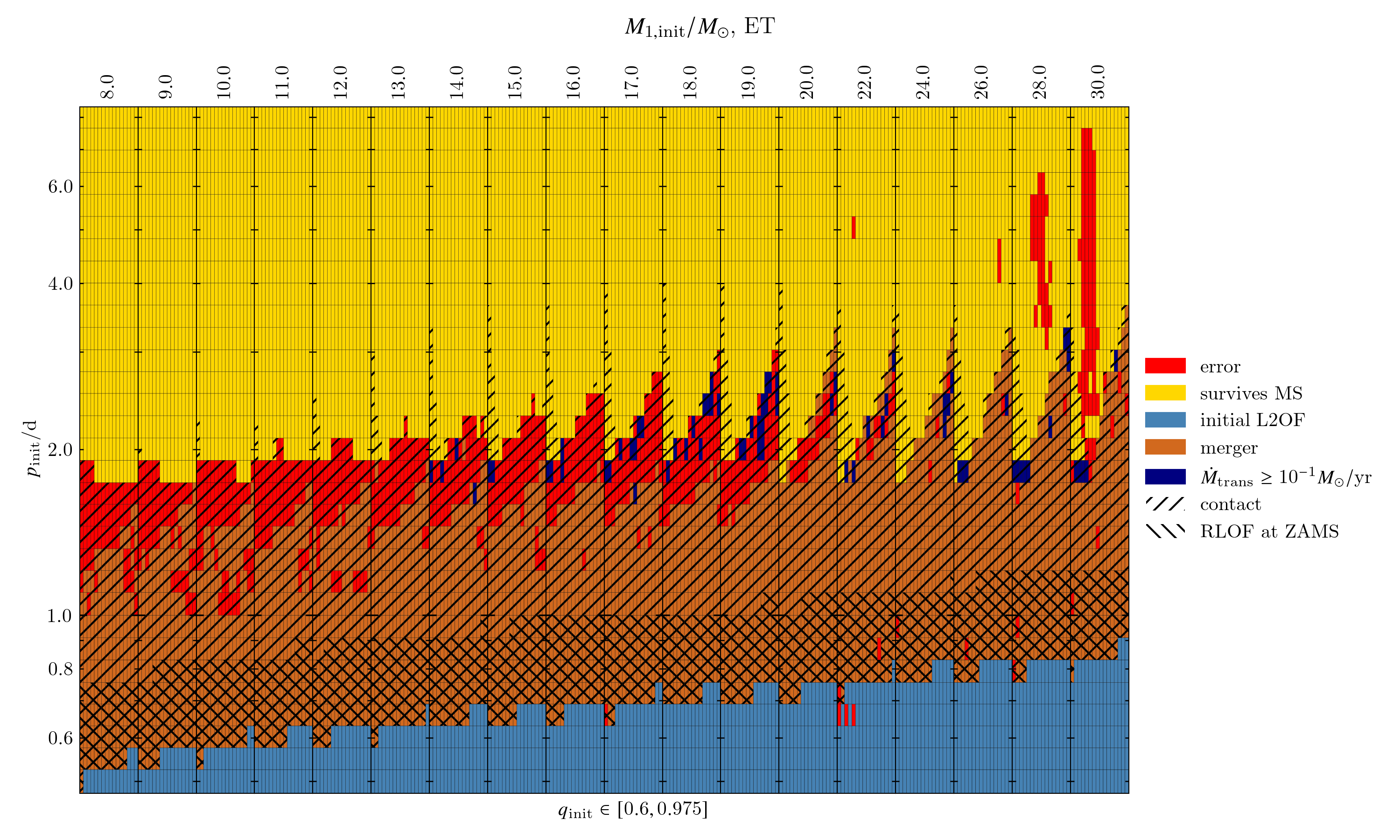}
	\caption{Outcomes of our grid of models that include ET.
		Each wider column contains the models for the initial primary mass $M_{1, \rm init}$ indicated on the top.
		The columns, in turn, are subdivided horizontally by initial mass ratio $q_{\rm init}$, increasing from left to right from 0.6 to 0.975, and vertically by initial period, $P_{\rm init}$. \textit{(Continued on next page.)}}
	\label{fig:et_completion}
\end{sidewaysfigure*}
\begin{sidewaysfigure*}
	\ContinuedFloat
	\centering
	\includegraphics[width=\textwidth]{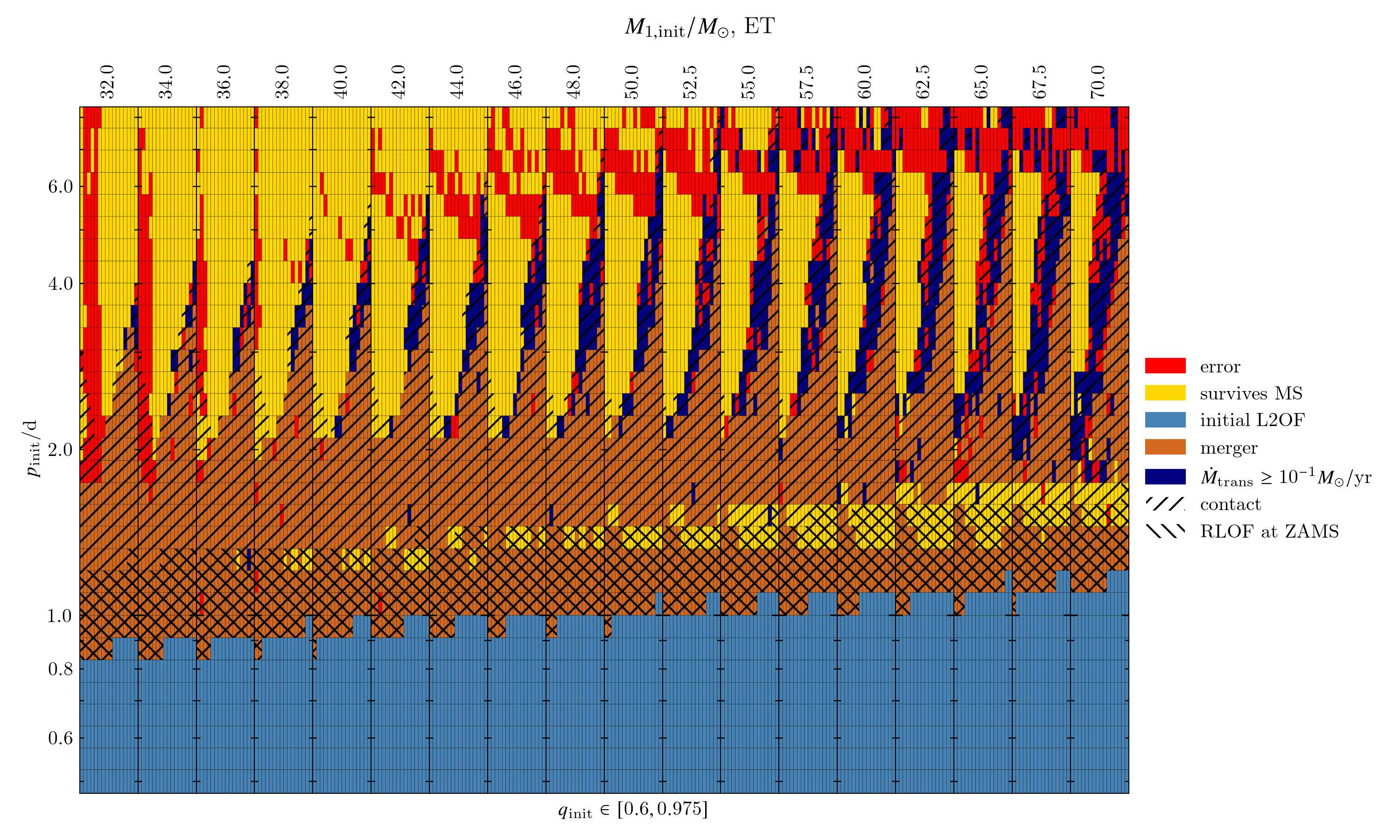}
	\caption[]{\textit{Continued.}}
\end{sidewaysfigure*}

\begin{sidewaysfigure*}
	\centering
	\includegraphics[width=\textwidth]{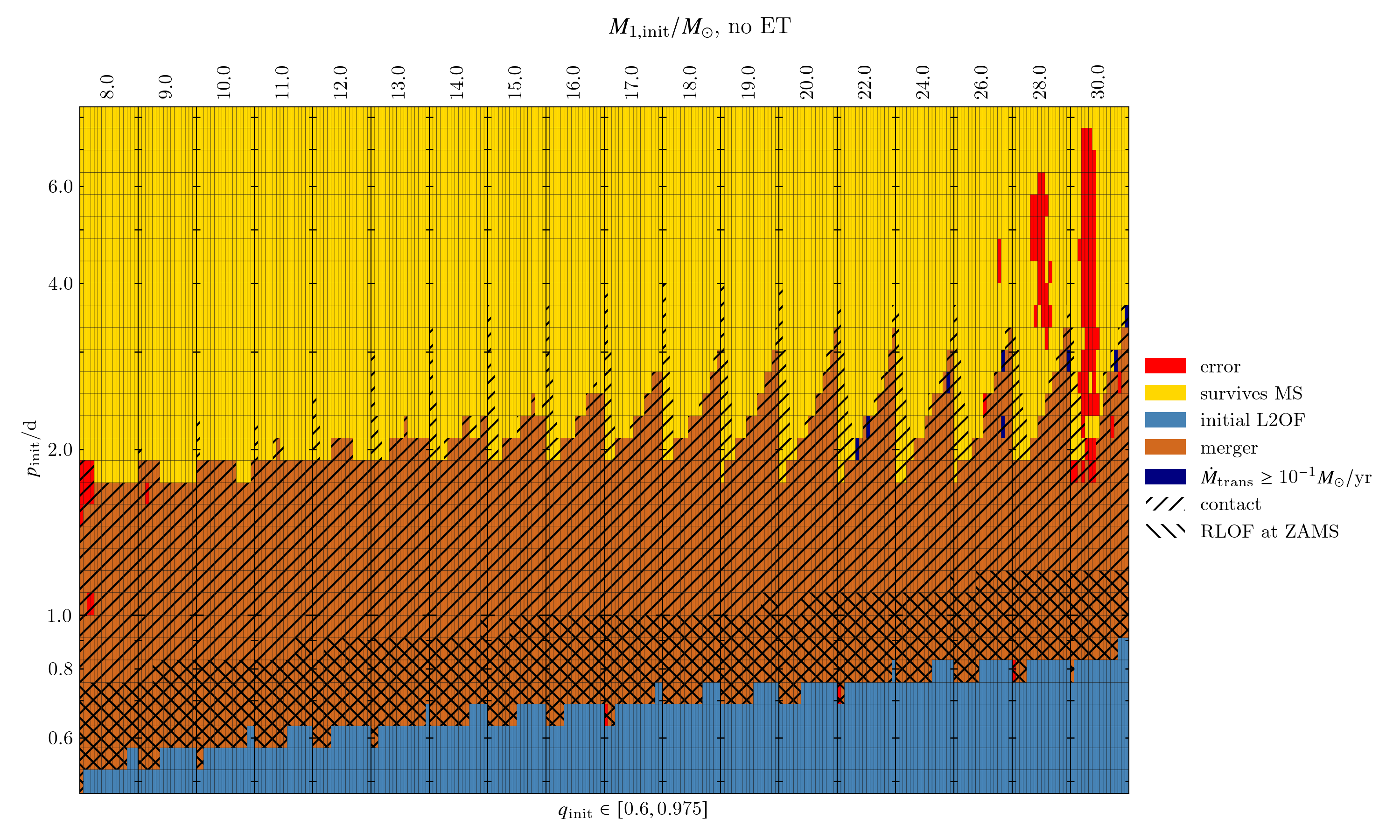}
	\caption{Outcomes of our grid of models that do not include ET. The axes and legend are similar to Fig.~\ref{fig:et_completion}. \textit{(Continued on next page.)}}
	\label{fig:noet_completion}
\end{sidewaysfigure*}
\begin{sidewaysfigure*}
	\ContinuedFloat
	\centering
	\includegraphics[width=\textwidth]{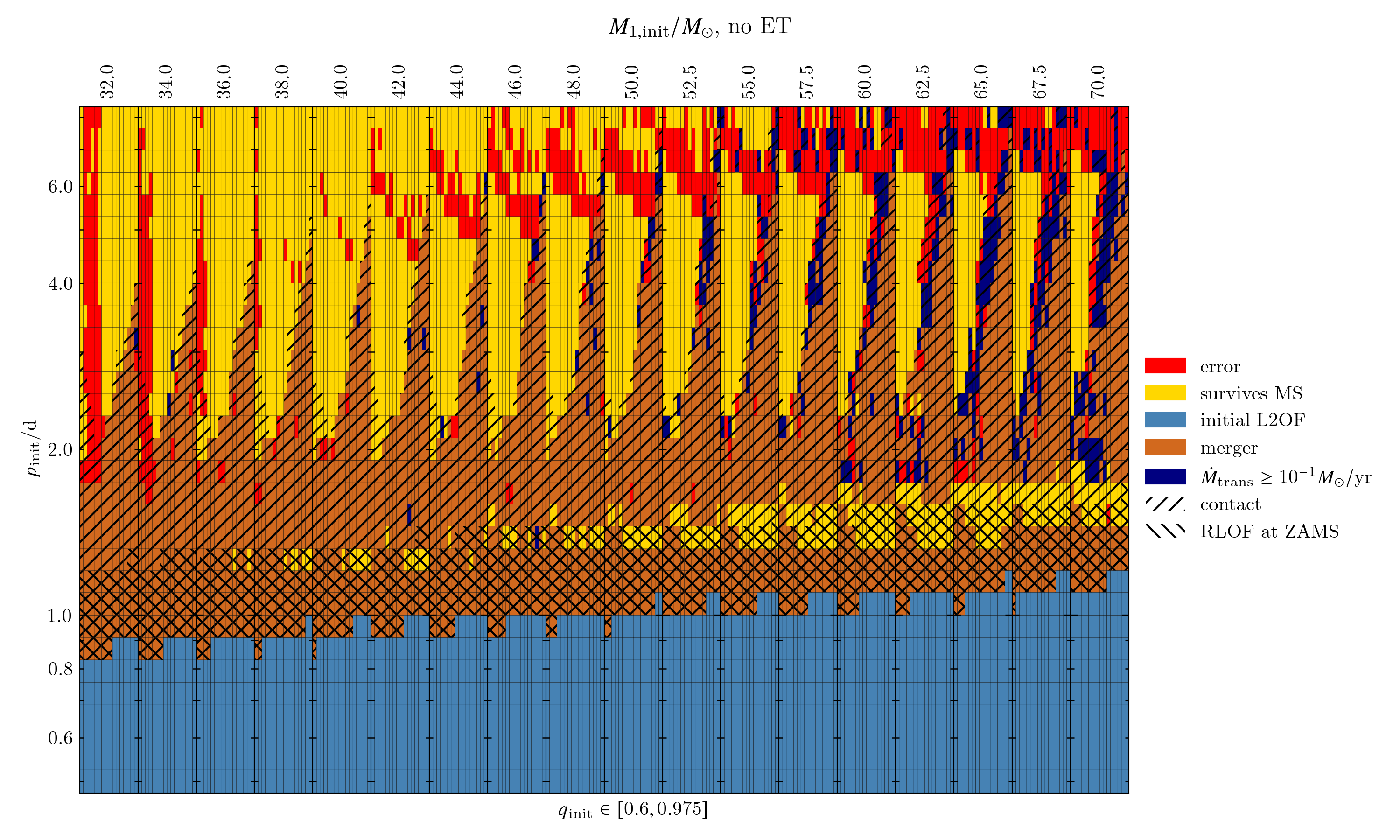}
	\caption[]{\textit{Continued.}}
\end{sidewaysfigure*}

\begin{sidewaysfigure*}
	\centering
	\includegraphics[width=\textwidth]{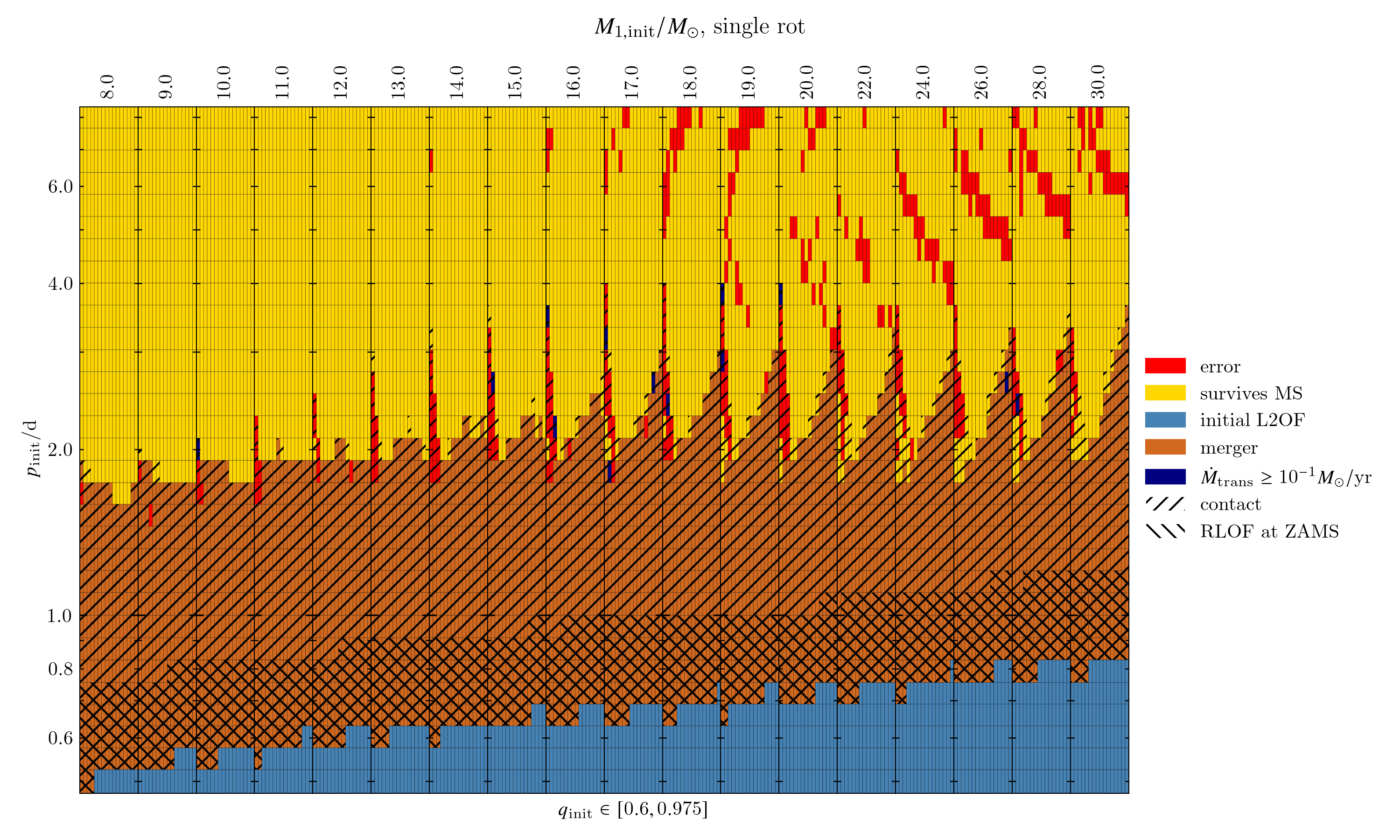}
	\caption{Outcomes of our grid of models that use the single-star rotation deformation. The axes and legend are similar to Fig.~\ref{fig:et_completion}. \textit{(Continued on next page.)}}
	\label{fig:single_rot_completion}
\end{sidewaysfigure*}
\begin{sidewaysfigure*}
	\ContinuedFloat
	\centering
	\includegraphics[width=\textwidth]{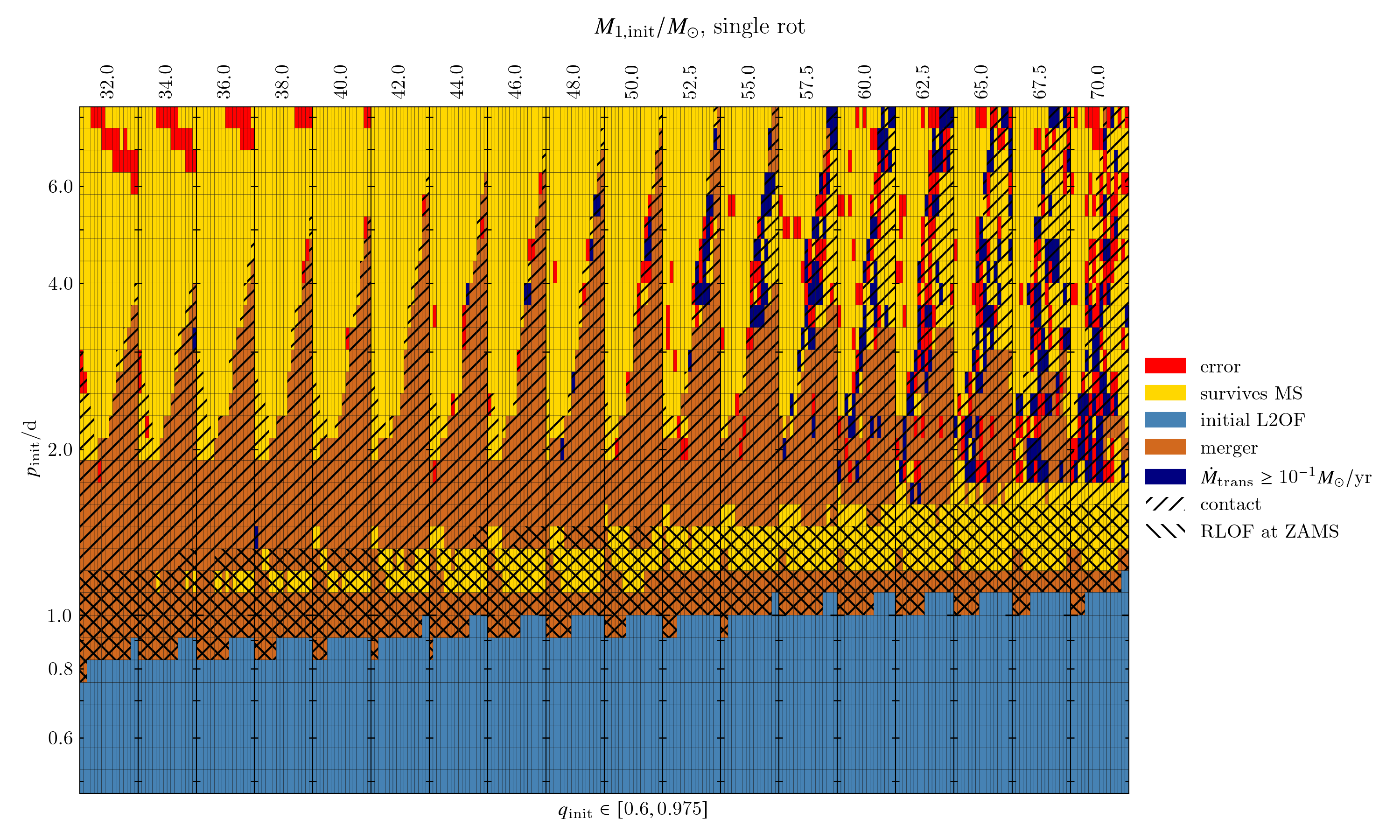}
	\caption[]{\textit{Continued.}}
\end{sidewaysfigure*}

\section{Observed Systems}
In this section, we present the parameters from observed massive (near-)contact binaries in the Milky way and M31 (Table \ref{tab:mw_systems}) and the Magellanic Clouds (Table \ref{tab:mc_systems}).
We took the period, mass ratio, masses, radii and temperatures from the indicated references, and computed $R/R_{\rm RL}$ according to the approximation of \citet{eggletonApproximationsRadiiRoche1983} and Kepler's third law.
We note that in the Tables of \citet{menonDetailedEvolutionaryModels2021}, some of their values, in particular on $R/R_{\rm RL}$, are misquoted.
\begin{table*}
\centering
	\begin{threeparttable}
		\caption{Parameters of observed, massive (near-)contact systems in the Milky Way and M31. For each system, the first (second) line gives the mass, radius, effective temperature and relative overflow for the primary (secondary).}
		\label{tab:mw_systems}
		\begin{tabular}{l c c c c c c l}
			\toprule\toprule
			Name & $p(\unit{\day})$ & $q = M_2/M_1$ & $M(\unit{\Msun})$ & $R(\unit{\Rsun})$ & $T_{\rm eff}(\unit{\kilo\kelvin})$ & $R/R_{\rm RL}$ & Reference \\
			\midrule
			V382 Cyg & $1.886$ & $\num{0.726(0.017)}$ & $\num{26.1(0.4)}$ & $\num{9.4(0.2)}$ & $\num{37.20(0.69:0.72)}$ & $\num{1.01}$ & A21 \\
			& & & $\num{19.0(0.3)}$ & $\num{8.7(0.2)}$ & $\num{38.25(0.73:0.75)}$ & $\num{1.08}$ & \\
			TU Mus & $1.387$ & $\num{0.625(0.009)}$ & $\num{16.7(0.4)}$ & $\num{7.2(0.5)}$ & $\num{38.7}$ & $\num{1.09}$ & P08\\
			& & & $\num{10.4(0.4)}$ & $\num{5.7(0.5)}$ & $\num{33.2}$ & $\num{1.06}$ & \\
			LY Aur & $4.002$ & $\num{0.550(0.007)}$ & $\num{25.5}$ & $\num{16.1}$ & $\num{31.0}$ & $\num{1.03}$ & Mr13 \\
			& & & $\num{14.0}$ & $\num{12.6}$ & $\num{31.2}$ & $\num{1.03}$ & \\
			V701 Sco & $0.762$ & $\num{0.995(0.002)}$ & $\num{9.78(0.22)}$ & $\num{4.137(0.316)}$ & $\num{23.5(1.0)}$ & $\num{1.14}$ & Y19\\
			& & & $\num{9.74(0.22)}$ & $\num{4.132(0.312)}$ & $\num{23.44(0.05)}$ & $\num{1.15}$ & \\
			CT Tau & $0.667$ & $\num{0.983(0.003)}$ & $\num{14.25(3.30)}$ & $\num{4.89(0.41)}$ & $\num{25.45(2.25)}$ & $\num{1.31}$ & Y19 \\
			& & & $\num{14.01(3.40)}$ & $\num{4.89(0.42)}$ & $\num{25.64(0.12)}$ & $\num{1.33}$ & \\
			GU Mon & $0.897$ & $\num{0.976(0.003)}$ & $\num{8.79(0.13)}$ & $\num{4.636(0.252)}$ & $\num{28.0(2.0)}$ & $\num{1.19}$ & Y19 \\
			& & & $\num{8.58(0.12)}$ & $\num{4.596(0.246)}$ & $\num{27.82(0.07)}$ & $\num{1.21}$ & \\
			XZ Cep & $5.097$ & $\num{0.50(0.02)}$ & $\num{18.7(1.3)}$ & $\num{14.2(0.1)}$ & $\num{28.0(1.0)}$ & $\num{0.85}$ & Ms17 \\
			& & & $\num{9.3(0.5)}$ & $\num{14.2(0.1)}$ & $\num{24.0(3.0)}$ & $\num{1.17}$ & \\
			LSS 3074 & $2.184$ & $\num{0.86(0.04)}$ & $\num{17.2(1.4)}$ & $\num{8.2(0.7)}$ & $\num{39.9(1.5)}$ & $\num{0.91}$ & R17 \\
			& & & $\num{14.8(1.1)}$ & $\num{7.5(0.6)}$ & $\num{34.1(1.5)}$ & $\num{0.93}$ & \\
			MY Cam & $1.175$ & $\num{0.839(0.027)}$ & $\num{37.7(1.6)}$ & $\num{7.60(0.10)}$ & $\num{42.0(1.5)}$ & $\num{1.00}$ & A21 \\
			& & & $\num{31.6(1.4)}$ & $\num{7.01(0.09)}$ & $\num{39.0(1.5)}$ & $\num{1.01}$ & \\
			V348 Car & $5.562$ & $\num{0.90(0.02)}$ & $\num{32(4)}$ & $\num{18.8(1.4)}$ & $\num{29.7(1.3)}$ & $\num{0.93}$ & HE85 \\
			& & & $\num{29(4)}$ & $\num{19.3(1.4)}$ & $\num{26.2}$ & $\num{1.00}$ & \\
			V729 Cyg & $6.598$ & $\num{0.290(0.012)}$ & $\num{31.6(2.9)}$ & $\num{25.6(1.1)}$ & $\num{28.0}$ & $\num{1.03}$ & YY14 \\
			& & & $\num{8.8(0.3)}$ & $\num{14.5(1.0)}$ & $\num{21.26(0.37)}$ & $\num{1.03}$ & \\
			BH Cen & $0.792$ & $\num{0.844(0.003)}$ & $\num{9.4(5.4)}$ & $\num{4.0(0.7)}$ & $\num{17.9}$ & $\num{1.09}$ & Lg84 \\
			& & & $\num{7.9(5.4)}$ & $\num{3.7(0.7)}$ & $\num{17.43(0.02)}$ & $\num{1.09}$ & \\
			SV Cen & $1.659$ & $\num{0.800(0.078)}$ & $\num{9.6}$ & $\num{7.8}$ & $\num{16}$ & $\num{1.01}$ & LS91 \\
			& & & $\num{7.7}$ & $\num{7.3}$ & $\num{24}$ & $\num{1.08}$ & \\
			V606 Cen & $1.495$ & $\num{0.548(0.001)}$ & $\num{14.7(0.4)}$ & $\num{6.83(0.06)}$ & $\num{29.2}$ & $\num{1.01}$ & Lz99 \\
			& & & $\num{7.96(0.22)}$ & $\num{5.19(0.05)}$ & $\num{21.77(0.02)}$ & $\num{1.08}$ & \\
			HD 64315 B & $1.019$ & $\num{1.00(0.09)}$ & $\num{14.6(2.3)}$ & $\num{5.52(0.55)}$ & $\num{32}$ & $\num{1.10}$ & Lo17 \\
			& & & $\num{14.6(2.3)}$ & $\num{5.33(0.52)}$ & $\num{32}$ & $\num{1.07}$ & \\
			V745 Cas & $1.411$ & $\num{0.571(0.010)}$ & $\num{18.31(0.51)}$ & $\num{6.94(0.07)}$ & $\num{30000}$ & $0.99$ & \c{C}a14 \\
			& & & $\num{10.47(0.28)}$ & $\num{5.35(0.05)}$ & $\num{25.54(0.30)}$ & $\num{0.99}$ & \\
			V4741 M31A & $1.604$ & $\num{0.924(0.015)}$ & $\num{18}^\dagger$ & $\num{7.2}^\dagger$ & $\num{31.6}$ & $\num{0.99}$ & Li22 \\
			& & & $\num{16.8}^\dagger$ & $\num{6.7}^\dagger$ & $\num{27.3}$ & $\num{0.97}$ & \\
			V1555 M31A & $0.917$ & $\num{0.974(0.012)}$ & $\num{23}^\dagger$ & $\num{6.7}^\dagger$ & $\num{35.1}$ & $\num{1.24}$ & Li22 \\
			& & & $\num{22.4}^\dagger$ & $\num{6.4}^\dagger$ & $\num{34.4}$ & $\num{1.24}$ & \\
			\bottomrule
		\end{tabular}
		\begin{tablenotes}
			\item [\textbf{References}] A21: \citet{abdul-masihConstrainingOvercontactPhase2021}; P08: \citet{pennyTomographicSeparationComposite2008}; Mr13: \citet{mayerOtypeEclipsingContact2013}; Y19: \citet{yangComprehensiveStudyThree2019}; Ms17: \citet{martinsPropertiesSixShortperiod2017}; R17: \citet{raucqObservationalSignaturesMassexchange2017}; HE85: \citet{hilditchMassiveNearcontactBinary1985}; YY14: \citet{yasarsoyInteractingSupergiantClose2014}; Lg84: \citet{leungRevisedUBVPhotometric1984}; LS91: \citet{linnellDoesSVCentauri1991}; Lz99: \citet{lorenzV606CentauriEarlytype1999}; Lo17: \citet{lorenzoMassiveMultipleSystem2017}; \c{C}a14: \citet{cakirliV745CassiopeanInteracting2014}, Li22: \citet{liTwoMassiveClose2022}
			\item [${}^\dagger$] $M_1$ is adopted, from which $M_2$ follows via $q$. The radii are found from the fractional radii from the photometric solution and Kepler's third law.
		\end{tablenotes}
	\end{threeparttable}
\end{table*}

\begin{table*}
\centering
    \begin{threeparttable}
    \caption{Parameters of observed, massive (near-)contact systems in the Magellanic Clouds. For each system, the first (second) line gives the mass, radius, effective temperature and relative overflow for the primary (secondary).}
    \label{tab:mc_systems}
    \begin{tabular}{l c c c c c c l}
    \toprule\toprule
    Name & $p(\unit{\day})$ & $q = M_2/M_1$ & $M(\unit{\Msun})$ & $R(\unit{\Rsun})$ & $T_{\rm eff}(\unit{\kilo\kelvin})$ & $R/R_{\rm RL}$ & Reference \\
    \midrule
    VFTS 352 & $1.124$ & $\num{0.99(0.01)}$ & $\num{28.85(0.30)}$ & $\num{7.25(0.02)}$ & $\num{41.45(1.17:0.80)}$ & $\num{1.01}$ & A21 \\
    & & & $\num{28.63(0.30)}$ & $\num{7.22(0.02)}$ & $\num{44.15(1.1:1.2)}$ & $\num{1.08}$ & \\
    VFTS 066 & $1.141$ & $\num{0.52(0.05)}$ & $\num{13.0(7.0:5.0)}$ & $\num{5.8(0.5:0.8)}$ & $\num{32.8(1.7:1.0)}$ & $\num{1.07}$ & M20 \\
    & & & $\num{6.6(3.5:2.8)}$ & $\num{4.4(0.4:0.8)}$ & $\num{29.0(1.0:1.2)}$ & $\num{1.09}$ & \\
    VFTS 661 & $1.266$ & $\num{0.71(0.02)}$ & $\num{27.3(0.9:1.0)}$ & $\num{6.80(0.04:0.01)}$ & $\num{38.4(0.9:0.4)}$ & $\num{0.94}$ & M20 \\
    & & & $\num{19.4(0.6:0.7)}$ & $\num{5.70(0.03:0.01)}$ & $\num{31.8(1.4:0.6)}$ & $\num{0.92}$ & \\
    VFTS 217 & $1.855$ & $\num{0.83(0.01)}$ & $\num{46.8(11.7)}$ & $\num{10.1(1.5:1.2)}$ & $\num{45.0(0.6:0.4)}$ & $\num{0.91}$ & M20 \\
    & & & $\num{38.9(9.7)}$ & $\num{9.4(1.4:1.0)}$ & $\num{41.8(1.7:0.6)}$ & $\num{0.92}$ & \\
    VFTS 563 & $1.217$ & $\num{0.76(0.01)}$ & $\num{26.2(11.9:5.2)}$ & $\num{6.6(0.4:0.7)}$ & $\num{32.4(1.0:0.9)}$ & $\num{0.95}$ & M20 \\
    & & & $\num{20.0(9.1:3.9)}$ & $\num{5.8(0.4:0.6)}$ & $\num{32.4(1.2:0.9)}$ & $\num{0.95}$ & \\
    MACHO CB${}^\dagger$ & $1.217$ & $\num{0.64(0.01)}$ & $\num{41.2(1.2)}$ & $\num{9.56(0.02)}$ & $\num{50}$ & $\num{1.06}$ & O01 \\
    & & & $\num{27.0(1.2)}$ & $\num{7.99(0.05)}$ & $\num{49.5}$ & $\num{1.09}$ & \\
    SMC 108086${}^\ddagger$ & $0.883$ & $\num{0.85(0.11)}$ & $\num{16.9(1.2)}$ & $\num{5.7(0.2)}$ & $\num{36.00(0.69:0.66)}$ & $\num{1.19}$ & A21 \\
    & & & $\num{14.3(1.7)}$ & $\num{5.3(0.2)}$ & $\num{35.20(0.64:0.72)}$ & $\num{1.19}$ & \\
    \bottomrule
    \end{tabular}
    \begin{tablenotes}
        \item [\textbf{References}] A21: \citet{abdul-masihConstrainingOvercontactPhase2021}; M20: \citet{mahyTarantulaMassiveBinary2020b}; O01: \citet{ostrovOrbitalSolutionMACHO2001}
        \item [${}^\dagger$] MACHO* J053441.3-693139
        \item [${}^\ddagger$] OGLE SMC-SC10 108086
    \end{tablenotes}
    \end{threeparttable}
\end{table*}

\FloatBarrier
 \end{appendix}
\end{document}